\documentclass[conference]{IEEEtran}
\makeatletter
\def\ps@headings{%
\def\@oddhead{\mbox{}\scriptsize\rightmark \hfil \thepage}%
\def\@evenhead{\scriptsize\thepage \hfil \leftmark\mbox{}}%
\def\@oddfoot{}%
\def\@evenfoot{}}
\makeatother
\IEEEoverridecommandlockouts
\usepackage{cite}
\usepackage{amsmath,amssymb,amsfonts}
\usepackage{algorithmic}
\usepackage{graphicx}
\usepackage{textcomp}
\usepackage{xcolor}

\usepackage{color}
\usepackage{url}
\usepackage[latin9]{inputenc}
\usepackage{amsthm}
\usepackage{bm}
\usepackage{amsthm}
\usepackage{graphicx}
\usepackage{epstopdf}
\usepackage{wrapfig}
\usepackage{algorithmic}
\usepackage{algorithm}
\usepackage{mathrsfs}
\usepackage{float}
\usepackage{mathtools}
\usepackage{tabulary}
\usepackage{booktabs}
\usepackage{caption}
\usepackage{subfig}
\usepackage{xprintlen}
\usepackage{multirow}
\usepackage[export]{adjustbox}

\def\BibTeX{{\rm B\kern-.05em{\sc i\kern-.025em b}\kern-.08em
    T\kern-.1667em\lower.7ex\hbox{E}\kern-.125emX}}

\graphicspath{{./}{./files/figures_eps/}}

\PassOptionsToPackage{bookmarks={false}}{hyperref} 

\newcommand{\comment}[1]{ }
\newcommand{\proposed}{{\tt{Tweak}}}

\newcommand\subparagraph{%
  \@startsection{subparagraph}{0}
  {\parindent}
  {0ex \@plus 0ex \@minus 0ex}
  {-1em}
  {\normalfont\normalsize\bfseries}}
\makeatother

\usepackage{titlesec}
\titlespacing*{\section}
{0pt}{1.2ex}{0ex} 
\titlespacing*{\subsection}
{0pt}{1ex}{0ex}  
\titlespacing*{\subsubsection}
{0pt}{0ex}{0ex} 

\begin{document}

\title{Tweak: Towards Portable Deep Learning Models for Domain-Agnostic LoRa Device Authentication
\thanks{This work is supported in part by Intel/NSF MLWiNS Award No. 2003273. ~\\
An IEEE-formatted version of this article is to appear in the 2022 IEEE Conference on Communications and Network Security (IEEE CNS 2022). Personal use of this material is permitted. Permission from IEEE must be obtained for all other uses, in any current or future media, including reprinting/republishing this material for advertising or promotional purposes, creating new collective works, for resale or redistribution to servers or lists, or reuse of any copyrighted component of this work in other works.}
}

\author{\IEEEauthorblockN{Jared Gaskin}
\IEEEauthorblockA{
\textit{Oregon State University}\\
Corvallis, Oregon, USA \\
gaskinj@oregonstate.edu}\vspace{-1cm}
\and
\IEEEauthorblockN{Bechir Hamdaoui}
\IEEEauthorblockA{
\textit{Oregon State University}\\
Corvallis, Oregon, USA \\
hamdaoui@oregonstate.edu}\vspace{-1cm}
\and
\IEEEauthorblockN{Weng-Keen Wong}
\IEEEauthorblockA{
\textit{Oregon State University}\\
Corvallis, Oregon, USA \\
wongwe@oregonstate.edu}\vspace{-1cm}}

\maketitle

\begin{abstract}
Deep learning based device fingerprinting has emerged as a key method of identifying and authenticating devices solely via their captured RF transmissions. Conventional approaches are not portable to different domains in that if a model is trained on data from one domain, it will not perform well on data from a different but related domain. Examples of such domains include  the receiver hardware used for collecting the data, the day/time on which data was captured, and the protocol configuration of devices. 
This work proposes \proposed, a technique that, using metric learning and a calibration process, enables a model trained with data from one domain to perform well on data from another domain. This process is accomplished with only a small amount of training data from the target domain and without changing the weights of the model, which makes the technique computationally lightweight and thus suitable for resource-limited IoT networks. This work evaluates the effectiveness of \proposed\ vis-a-vis its ability to identify IoT devices using a testbed of real LoRa-enabled devices under various scenarios. The results of this evaluation show that \proposed\ is viable and especially useful for networks with limited computational resources and applications with time-sensitive missions.
\end{abstract}

\begin{IEEEkeywords}
Device authentication, domain-agnostic portable device fingerprints, learning model calibration.
\end{IEEEkeywords}

\section{Introduction}
\label{sec:into}
Recent years have seen unprecedented growth in both the number and variety of IoT networks and applications~\cite{IoTAnalytics-22}. Due to its ability for enabling long-range connectivity between IoT devices at low power, LoRa technology~\cite{LoRa-allicance} has been widely adopted by hundreds of IoT application developers as the de facto wide-area wireless network access protocol. With such a rapid adoption of LoRa technology coupled with the massive numbers of emerging resource-constrained IoT devices, there is undoubtedly an urgent need for {\em lightweight} and {\em scalable} authentication mechanisms that can ensure automated and secure access to these LoRa-enabled IoT networks.

Device authentication mechanisms that are based on RF (radio frequency) fingerprinting have recently been recognized as key methods with great potential for complementing conventional cryptographic approaches to increase the security protection of IoT networks against unauthorized access~\cite{eleven_domainGeneralization2022}. These fingerprinting methods essentially consist of extracting device-specific features (aka fingerprints) from received RF signals---typically caused by inevitable transceiver hardware impairments incurred during manufacturing, and leveraging these features to uniquely identify and classify wireless transmitters~\cite{four_scalableLoRa2022,eleven_domainGeneralization2022}.
Various feature extraction approaches have been taken, including early hand-crafted approaches that require RF signal domain knowledge and many trial-and-error attempts (e.g.~\cite{li2020wireless}), and more recent approaches that leverage deep learning  to extract features automatically from raw RF signals without requiring RF domain expertise (e.g.~\cite{eleven_domainGeneralization2022,deployment_variability}). 

\subsection{Multi-Domain Portability Challenges}
\begin{figure}
\centering
\subfloat[Models trained on RX1 data]{
   \includegraphics[width=0.42\columnwidth]{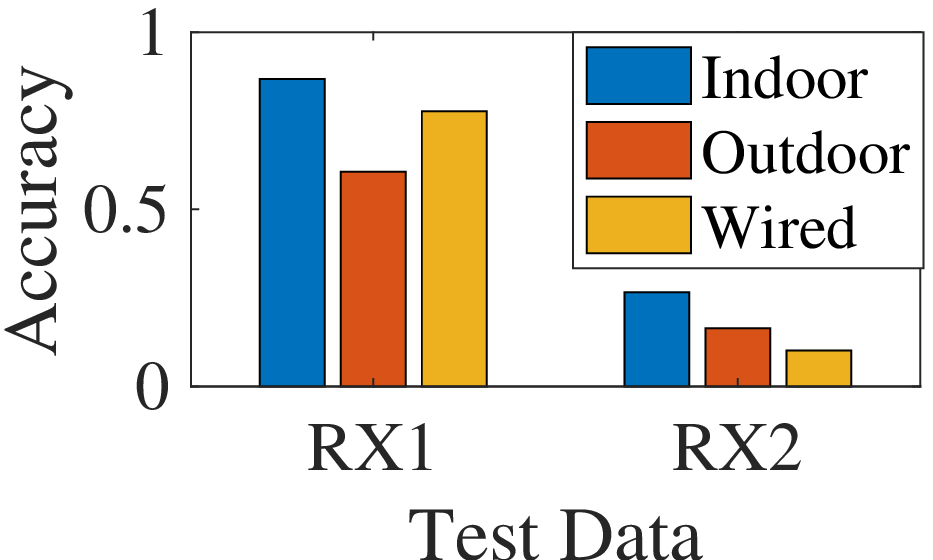}\label{subfig:introRX}}
\subfloat[Models trained on Day 1 data]{
   \includegraphics[width=0.49\columnwidth]{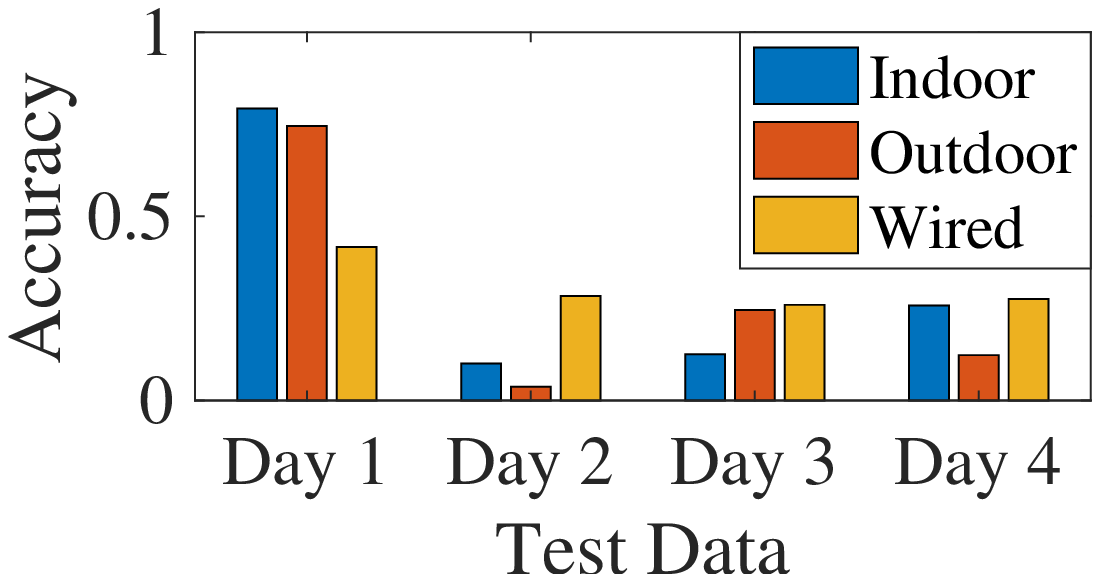}\label{subfig:introDay}}
  
   \comment{
\subfloat[Model trained on Config. 1 data]{
   \includegraphics[width=0.6\columnwidth]{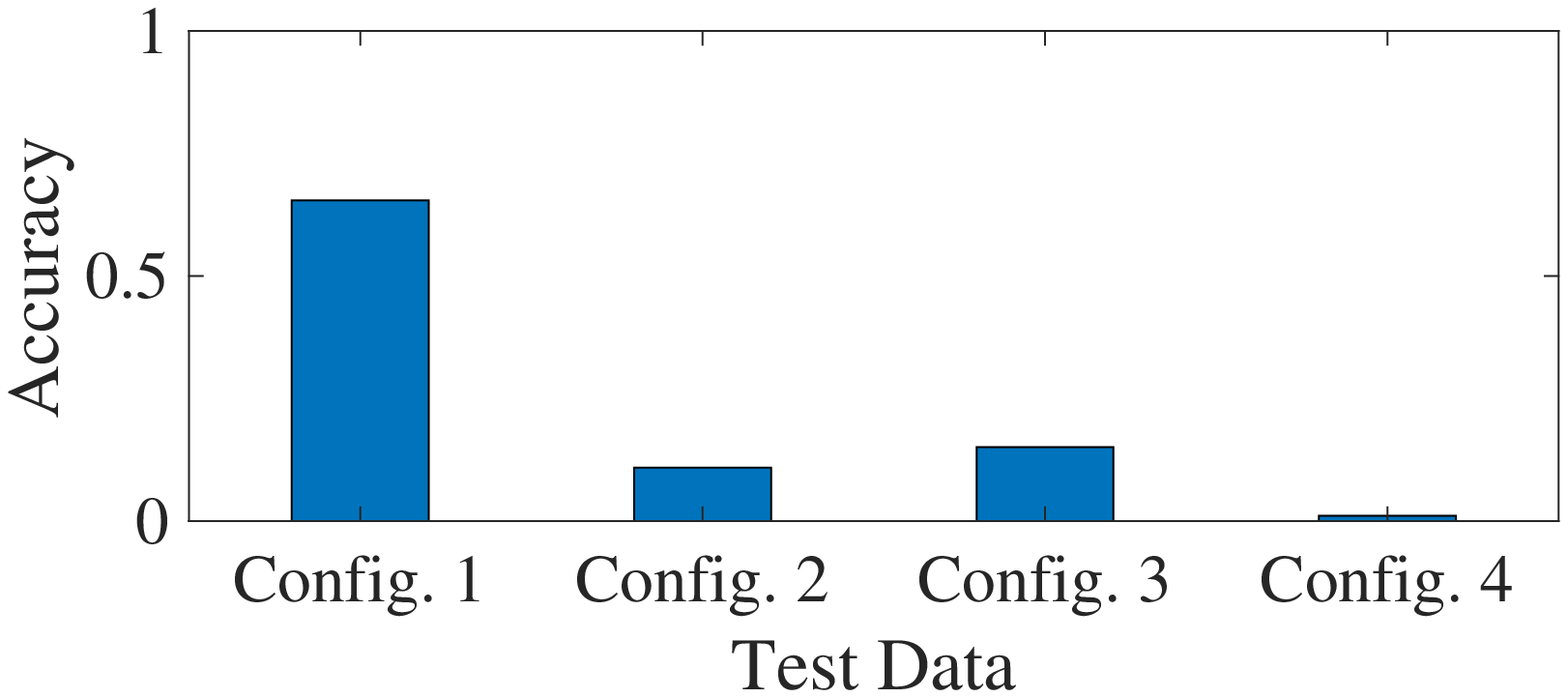}\label{subfig:introConfig}}
   }
\caption{Accuracy of models tested on different domains.}
\label{fig:intro_diff_day_diff_config_diff_rx}
\vspace{-.2in}
\end{figure}

Although recent deep learning approaches have shown promising results, they fail to maintain good performance when the data used during training and that used during testing are collected under different domains. For example, when the model is trained on data collected at one receiver but tested on data collected at a different receiver, the device identification accuracy degrades substantially compared to when both training and testing are done on data collected using the same receiver. 
%
To demonstrate the impact of such challenges, we performed several experiments using data collected with an IoT testbed, consisting of 25 LoRa transmitters and 2 USRP B210 receivers, under three different deployment settings: indoors, outdoors and wired (detailed description of the testbed and experimental scenarios is given later in Sec.~\ref{sec:setup}). 
Our results depicted in Fig.~\ref{subfig:introRX} show the accuracy of the learning model when trained on data captured at one receiver (RX1), but tested on data captured at a different receiver (RX2), as well as at the same receiver (RX1). These results clearly show the substantial drop in accuracy due to the change in the receiver hardware. As depicted in Fig.~\ref{subfig:introDay}, similar trends are also observed when the model is trained on data captured on Day 1 but tested on data captured on a different day (while keeping the same receiver). The common trend in these figures is that the models only perform well when the testing domain (receiver/day) matches the training domain.

The issues of domain portability we just illustrated have also been demonstrated in other works~\cite{deployment_variability,alshawabkainfocom,one_DeepRadioID2019}. Since this problem is well established, developing approaches that can overcome such issues has become an active research area in recent years.

\subsection{Limitations of Related Work}
Past works attempting to address the wireless channel portability problem ~\cite{one_DeepRadioID2019,two_deepLoRa2021,nine_practicalTraining2021,eleven_domainGeneralization2022} are almost all based on using data augmentation and/or custom feature extraction as preprocsessing in order to extract RF fingerprints that are channel agnostic. The downside of these approaches is that the resulting model is not flexible and must be re-trained if it cannot perform well in a new condition. A notable exception to this is the work in~\cite{one_DeepRadioID2019}, which proposed a system that can estimate the channel and adapt to it.

Other approaches~\cite{three_receiverAgnostic2019,six_addressPortability2018,eight_AdaptorWifiID2021} attempt to ensure portability with regard to receiver hardware. However, these methods require some form of potentially computationally intensive training or optimization in order to transform input data so that it can be correctly classified. 

The authors of~\cite{four_scalableLoRa2022} make perhaps the most notable recent attempt at portability, and employ a method similar to the one described herein in the sense that both methods use metric learning and the triplet loss function. They address both channel portability and transmitter hardware portability (training and testing using two different sets of transmitters). Notably, this work is the only one, to the best of our knowledge, that performs its evaluation on open-set identification and classification \cite{Scheirer13}, which allows inputs from unknown transmitters and attempts to reject these inputs.

The costly alternatives available to practitioners while research on portability continues are as follows: (i) training a new deep learning model for each domain one wishes to test with, and (ii) training a single deep learning model with data from multiple domains in an attempt to produce a domain-agnostic model. These approaches are both costly in terms of the extra data required from additional target domains and the computation time needed to train the neural network models.

\subsection{Our Contributions: \proposed}
\begin{figure}
\centering
\subfloat[Enabling receiver hardware portability]{
   \includegraphics[width=.9\columnwidth]{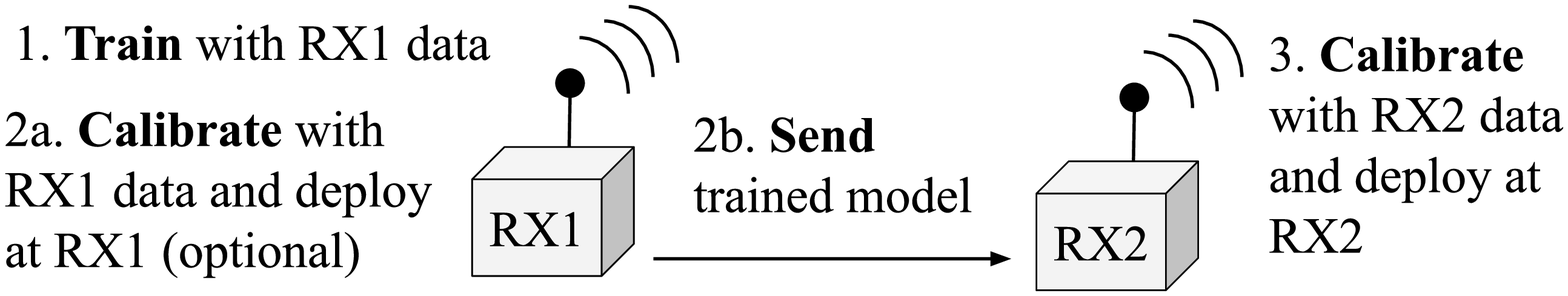}
   \label{fig:use_RX}}\\
\subfloat[Enabling wireless channel portability]{
   \includegraphics[width=.9\columnwidth]{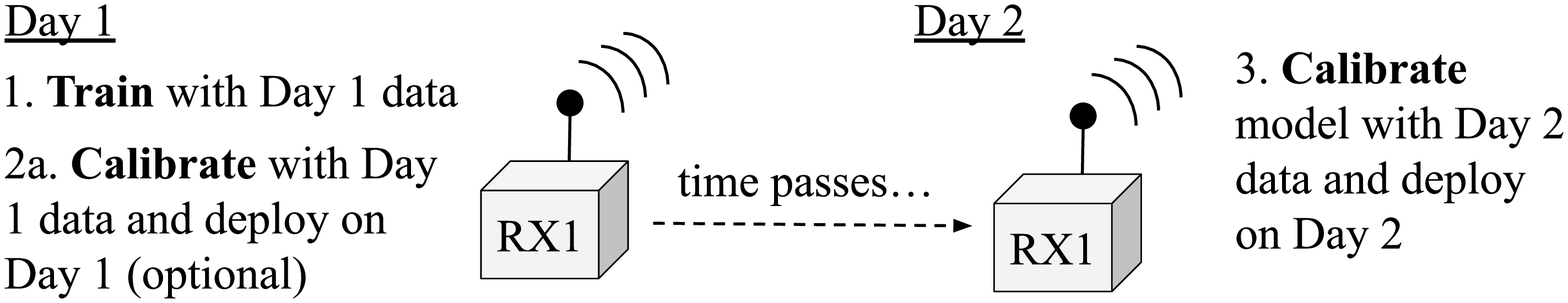}
   \label{fig:use_days}}\\
\subfloat[Enabling protocol configuration portability]{
   \includegraphics[width=.9\columnwidth]{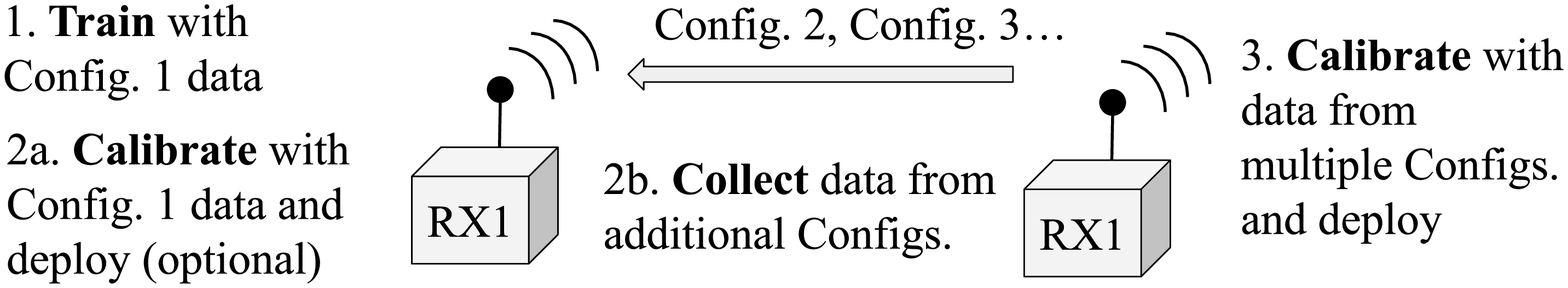}\label{subfig:M_config}
   \label{fig:use_configs}}
\caption{\proposed: Transferring of a trained model to a different domain: (a) hardware, (b) channel, and (c) configuration.}
\label{fig:use-cases}
\end{figure}

We propose \proposed, a lightweight device identification technique that enables portability of the deep learning models across multiple different domains. \proposed\ leverages metric learning to achieve accurate identification and open-set accuracy through model calibration instead of re-training, thus significantly reducing (i) the training time and (ii) the amount of needed data, making it more suitable for resource-constrained and real-time IoT applications. \proposed\ is evaluated using datasets collected from a testbed of 25 LoRa transmitters and 2 USRP B210 SDR receivers.
Specifically, \proposed\ enables:

\begin{itemize}
\item \textbf{Lightweight model portability} 
through calibration that can be performed very quickly, without changing the model weights, and can be done using a relatively small amount of labeled data from the target domain.

 \item \textbf{Multi-domain model portability} across multiple different domains, including receiver hardware, communication channel, and LoRa configuration, as shown in Fig.~\ref{fig:use-cases}.

\item \textbf{Open-set device identification} by easily acting as an open-set classifier, as \proposed\ handles inputs from unknown devices. Most related approaches that attempt to address portability do not perform open-set testing.

\end{itemize}

The remainder of this work is organized as follows: Sec.~\ref{sec:proposed} presents \proposed. Sec.~\ref{sec:setup} and~Sec.~\ref{sec:performance} evaluate \proposed\ using RF data captured using a real testbed of IoT devices. 

\section{\proposed: Enabling Deep Learning Model Portability for Robust Device Authentication}
\label{sec:proposed}

Our goal is to provide a device authentication method by which an open-set deep learning model can be trained on data from one domain (collected at one receiver, through one wireless channel, or with IoT transmitters using one LoRa configuration) and then be calibrated to perform well on data from a different but related domain. In this way, \proposed\ aims to move toward the creation of deep learning models for RF fingerprint authentication that are portable in terms of hardware, wireless channel, and LoRa protocol configuration.

\proposed\ achieves these portability goals through a calibration process that is: (i) not computationally intensive (and can be done on IoT hardware that is less powerful than the resources used for training the original model) and (ii) accomplished with a limited amount of labeled training data from the target domain (receiver, day, configuration).

\begin{figure} 
\centerline{
   {\includegraphics[width=.75\columnwidth]{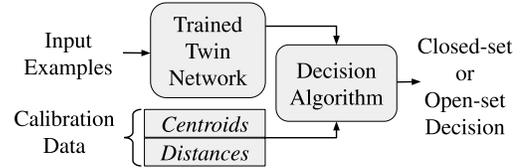}}}
   \vspace{1mm}
\caption{A high-level diagram of \proposed}    
\label{fig:proposed}
\end{figure}

Fig.~\ref{fig:proposed} summarizes \proposed.  Prior to deployment, \proposed\ requires a training phase for the twin neural network and a calibration process using data from the target domain from all devices. Note that the calibration process can be repeated with different data without altering the twin neural network.  The following subsections provide more information including background material on the twin network architecture, details regarding how the neural network is trained as well as specifics of the calibration and decision-making algorithms.

\subsection{Deep Learning}
This section focuses the discussion on metric learning and twin neural networks; a comprehensive treatment on deep learning is outside the scope of this work and additional information can be found in~\cite{Goodfellow-et-al-2016}.  \proposed\ uses twin neural networks as open-set classifiers. While closed-set classifiers are designed to classify data with the assumption that they only encounter data instances from the classes observed during training, open-set classifiers can handle data from classes (i.e., devices) that are not seen during training~\cite{Scheirer13}. We deal with data instances from these unseen classes by identifying them as not belonging to one of the known classes seen during training. While the open-set problem is more difficult, it is also more realistic and appropriate for device authentication when unknown devices are often encountered in deployment.

\subsubsection{Metric Learning and Twin Neural Networks}
The standard approach for building a deep neural network (DNN) classifier is to train the model using a cross entropy loss with one output node for each class. The values of these output nodes are then used to determine the degree to which an input instance is predicted to belong to a particular class. Under this standard supervised learning setting, it is non-trivial to adapt the network to different domains such as a different receiver, day or configuration. Sophisticated domain adaptation methods for deep learning (e.g. \cite{wilson2020survey,wang2018survey}) have been developed, but these methods are typically computationally intensive and require large amounts of data. In contrast, we want a lightweight domain adaptation process that can be quickly calibrated using a small number of labeled examples from the target domain.

To overcome these issues, our work makes use of deep metric learning \cite{ghojogh_spectral_2022} in which a DNN produces an embedding of a data instance in a $K$-dimensional latent space. Within this embedding space, data instances that are similar (e.g. from the same device) will be closer than those that are not similar (e.g. from different devices). \proposed\ uses a well known metric learning structure called a twin (or Siamese) neural network \cite{bromley1994,chopra2005}. These types of neural networks have been successfully used for applications where few examples are available such as signature verification~\cite{bromley1994} and facial similarity calculations~\cite{facialSiamese}. Additionally, it has been shown that twin networks trained on one image dataset (e.g. Omniglot) and evaluated on an entirely different image dataset (e.g. MNIST) are able to generalize and maintain some level of performance~\cite{Koch2015SiameseNN}. However, they have only been applied directly to the problem of device fingerprinting based on RF signals in one other work~\cite{four_scalableLoRa2022}. 

\begin{figure}
\centerline{
   {\includegraphics[width=.9\columnwidth]{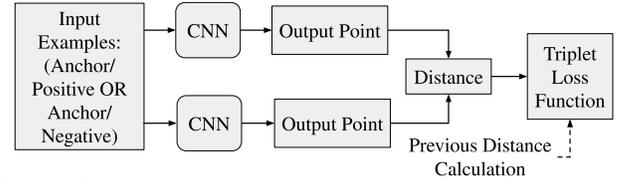}}}
  \vspace{-.07in}
\caption{General twin network architecture during training.}
\label{fig:siamese}
\end{figure}

Fig.~\ref{fig:siamese} depicts a twin neural network, consisting of a pair of convolutional neural networks (CNNs) constrained to have identical weights. Each CNN accepts a single input instance and produces an output that corresponds to coordinates in a $K$-dimensional latent embedding space ($K$ being the number of output nodes). The distance between the two outputs in the embedding space is computed as a measure of similarity and can be used to train the network with a loss function. In practice, a trained twin network can be considered as a single DNN that produces an output point in the embedding space, instead of twin networks that produce a distance (since the networks have identical weights). Hence, using a twin network does not impose additional memory costs.

\begin{figure}
\centering
\begin{tabular}{c|c}
\subfloat[Zero loss]{
   \includegraphics[width=0.4\columnwidth]{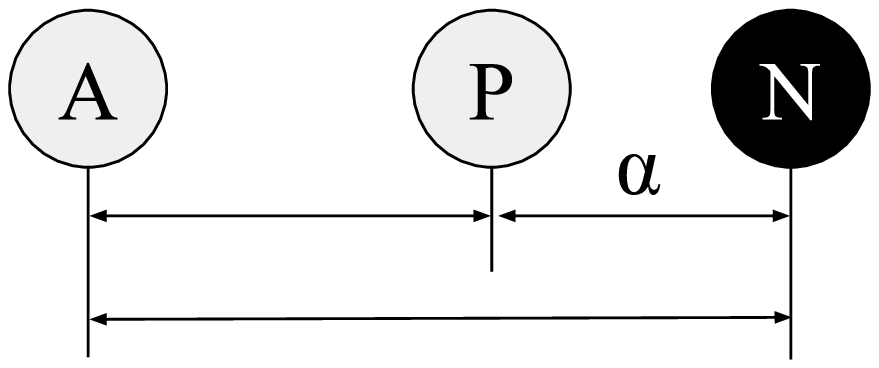}} &
\subfloat[Positive loss]{
   \includegraphics[width=0.4\columnwidth]{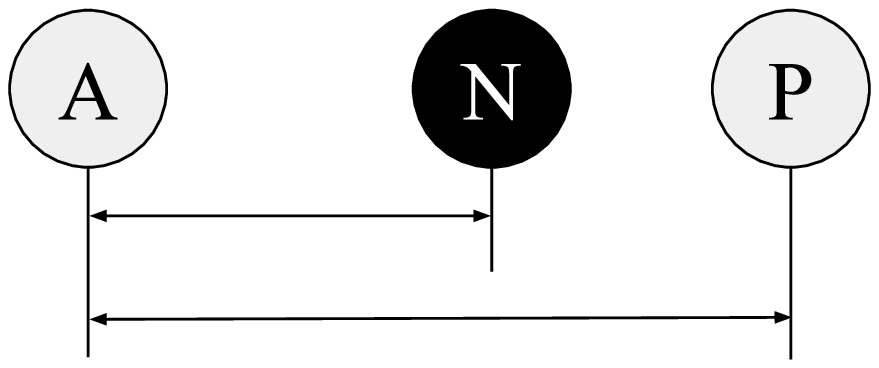}}
\end{tabular}
 \vspace{-.07in}
\caption{Example triplet loss values for one triplet.}    
\label{fig:triplet}
\end{figure}

\subsubsection{Training the Twin Neural Network}
Training a twin neural network requires a loss function that will cause data instances from different classes to be further apart in the output space and data instances from the same class to be closer together. A loss function that fits this criterion is the triplet loss function~\cite{tripletLoss}, which encourages reduced distances between `Anchor' (A) and `Positive' (P) examples, and increased distances between `Anchor' (A) and `Negative' (N) examples. The mathematical expression for this loss function is provided in Equation~\eqref{loss} and a visual depiction is provided in Fig.~\ref{fig:triplet}, where $\alpha$ is a margin value (indicating that the distance between P and N has to be at least this much to matter), and $f(\cdot)$ is the neural network mapping function.
\begin{equation}\small
    Loss = \max(\lVert f(A) - f(P) \rVert^2 - \lVert f(A) - f(N)\rVert^2 +\alpha,0)
    \label{loss}
\end{equation}

Triplets of A, P, and N examples are provided to the network during training. We generate the triplets after examples from a mini-batch have already passed through the network but before the loss is calculated. Determining triplets after examples have passed through the network is advantageous because more difficult triplets can be strategically selected to help with training~\cite{tripletLoss}, where difficult triplets are those with high loss due to having $A$ closer to $N$ than to $P$. 

\proposed\ uses the triplet loss function to train the twin network as shown in Fig.~\ref{fig:proposed}. Once this model has been trained on labeled data from a chosen domain, inputs from different transmitters should be mapped to outputs that are farther apart, and inputs from the same transmitter should be mapped to outputs that are closer together. At this point, the model is typically only capable of making decisions given two inputs that it can compare in terms of their distance in the latent space. To enable the trained twin network to make decisions about a single input from a particular domain, a calibration process must be performed. 

\subsection{Calibration and Decision Making Algorithms}

\begin{figure}
\begin{center}
   \includegraphics[width=.85\columnwidth]{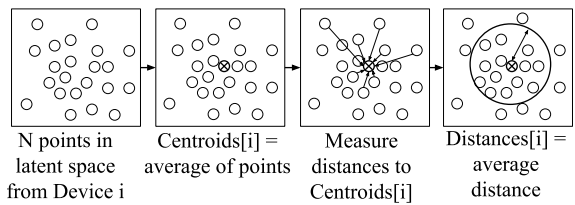}
\end{center}
\caption{Calibration for a single device, after $N$ examples from Device $i$ have passed through the trained twin network.}    
\label{fig:calibration}
\end{figure}

\begin{algorithm}\small
\caption{Calibration}\label{alg:calibration}
\begin{algorithmic}
\STATE $\bm{Centroids} = [\ ]$
\STATE $\bm{Distances} = [\ ]$
\STATE $\bm{Devices} = \{ \bm{D_1},\bm{D_2},...,\bm{D_{K}}$\} each $\bm{D_i} = \{\bm{x_{i1}},\bm{x_{i2}},...,\bm{x_{iN}}\}$
\FORALL{$\bm{D_i} \in \bm{Devices}$}
    \STATE $\bm{Centroids[i]} = {\sum_{j=1}^{N} f(\bm{x_{ij}})}/{N}$
    \STATE $\bm{Distances[i]} = {\sum_{j=1}^{N} \lVert f(\bm{x_{ij}}) - \bm{Centroids[i]} \rVert }/{N}$
\ENDFOR
\STATE $Calibrated \ for \ \bm{D_i} \in \bm{Devices}$

\end{algorithmic}
\end{algorithm}

The calibration process is performed by providing the trained network with labeled examples from all desired devices. Note that these examples may come from the training data if calibrating for the same domain as the training data. Algorithm~\ref{alg:calibration} describes the calibration process and Fig.~\ref{fig:calibration} illustrates it. We use the convention that boldface symbols refer to vectors and sets while non-boldface refer to scalars. For each device (class), we repeat the following process: given $N$ examples from device $i$ for calibration, $\bm{x_{i1}},\bm{x_{i2}},...,\bm{x_{iN}}$, we first pass these examples through the trained neural network, $f(\cdot)$, to produce $N$ corresponding outputs in the latent space. The first image in Fig.~\ref{fig:calibration} shows the result of this step. Next, these $N$ outputs are averaged to produce $\bm{Centroids[i]}$ (i.e. the centroid for device $i$), which is shown in the second image of Fig.~\ref{fig:calibration}. This centroid is the first piece of calibration data that will be stored for device $i$.

Next, the distances between the centroid for device $i$ and each of the $N$ outputs that were used to generate the centroid are calculated. This is depicted in the third image of Fig.~\ref{fig:calibration}. These distances are averaged to produce $\bm{Distances[i]}$ (i.e. the average distance for device $i$), which is represented by a circle with this distance value as its radius in Fig.~\ref{fig:calibration}. This distance is the second piece of calibration data that will be stored for device $i$. After the above process is repeated for every device, calibration is complete and the $\bm{Centroids}$ and $\bm{Distances}$ for the devices can be used to make decisions regarding new inputs using the decision algorithm to be described later.

This calibration process is designed to be lightweight as it only requires feeding data instances forward through the network and performing distance calculations, thus avoiding re-training the network. Additionally, the parameter $N$ represents the number of examples from each device used for calibration purposes. Altering $N$ allows adjusting the amount of computation required for calibration as well as the amount of labeled data required. The calibration process could also be repeated with data from multiple domains for the same devices in order to produce a model capable of making decisions about data from these multiple domains, thereby producing multiple $\bm{Centroids}$ and $\bm{Distances}$ for each device.

\begin{algorithm}\small
\caption{Open-set Binary Decision}
\label{alg:decisionOpen}
\begin{algorithmic}
\STATE $\bm{InputDeviceExamples} = \{\bm{x_1},\bm{x_2},...,\bm{x_M}\}$
\STATE $\bm{InputPoint} = {\sum_{j=1}^{M} f(\bm{x_{j}})}/{M}$
\FORALL{$\bm{D_i} \in \bm{Devices}$}
    \IF{$dist(\bm{InputPoint}, \bm{Centroids[i]}) \leq \bm{Distances[i]}$}
        \STATE Return: $Decision = \underline{\textbf{Admit}}$
    \ENDIF
\ENDFOR
\STATE Return: $Decision = \underline{\textbf{Reject}}$

\end{algorithmic}
\end{algorithm}
Once calibrated, the network is then used as an open-set classifier (Algorithm~\ref{alg:decisionOpen}) to classify an input data instance as belonging to a known or unknown device. The algorithm begins by collecting $M$ examples, $\bm{x_1},\bm{x_2},...,\bm{x_M}$, from an input device, passing these examples through the trained network, $f(\cdot)$, and averaging the $M$ outputs to produce $\bm{InputPoint}$.

Since this operation is not typical in deep learning models, it deserves some explanation. In the RF domain, the input data represents fractions of a second of real-time RF communication. Thus, it is plausible to collect $M$ input examples quickly in order for the model to process them for the open-set classifier. Doing so allows for the construction of a less noisy $\bm{InputPoint}$ that is more representative of a particular device and less susceptible to the effects of outlier examples.

After $\bm{InputPoint}$ is computed, for each device $i$ the algorithm calculates the distance between the $\bm{InputPoint}$ and $\bm{Centroids[i]}$ and compares this distance with $\bm{Distances[i]}$. If this calculated distance is less than or equal to $\bm{Distances[i]}$, the algorithm decides that the input comes from a known device, and returns the `Admit' decision (i.e., device is authenticated). If none of the calculated distances are less than or equal to $\bm{Distances[i]}$, then the algorithm decides the input came from an unknown device, and returns the `Reject' decision (i.e., device is denied network access).

\section{Experimental Scenarios and Datasets}
\label{sec:setup}

\begin{figure}
\centering
\begin{tabular}{m{0.3\columnwidth}|m{0.7\columnwidth}}
\multirow{2}{*}[5em]{
\subfloat[]{
   \includegraphics[height=2.9in]{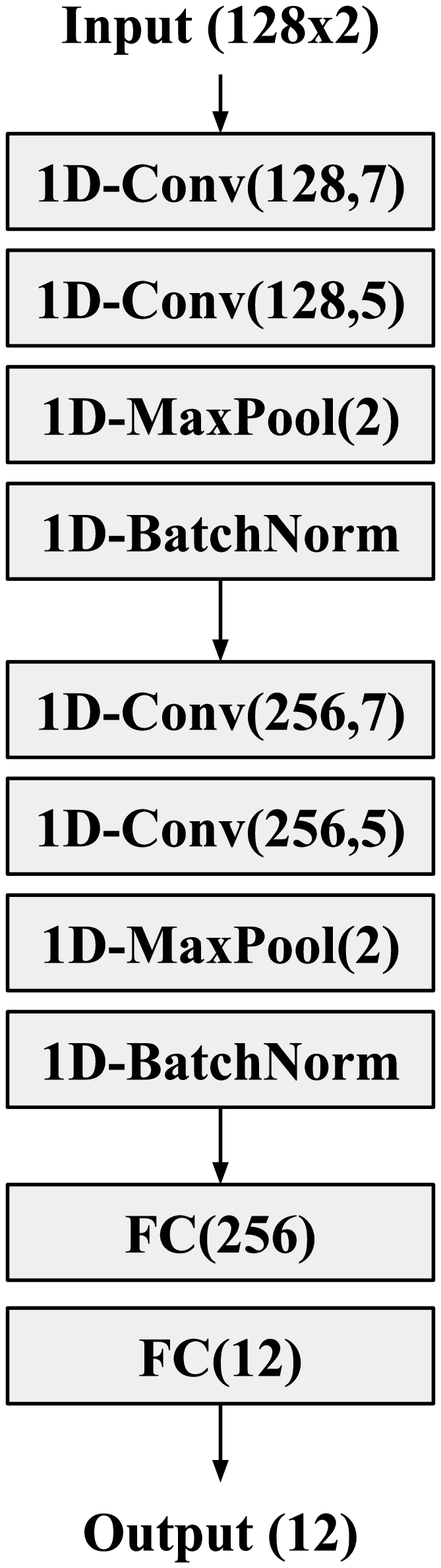}\label{subfig:twinNet}}}
&
\subfloat[USRP B210 SDRs (RX1 and RX2)]{
   \includegraphics[width=0.5\columnwidth]{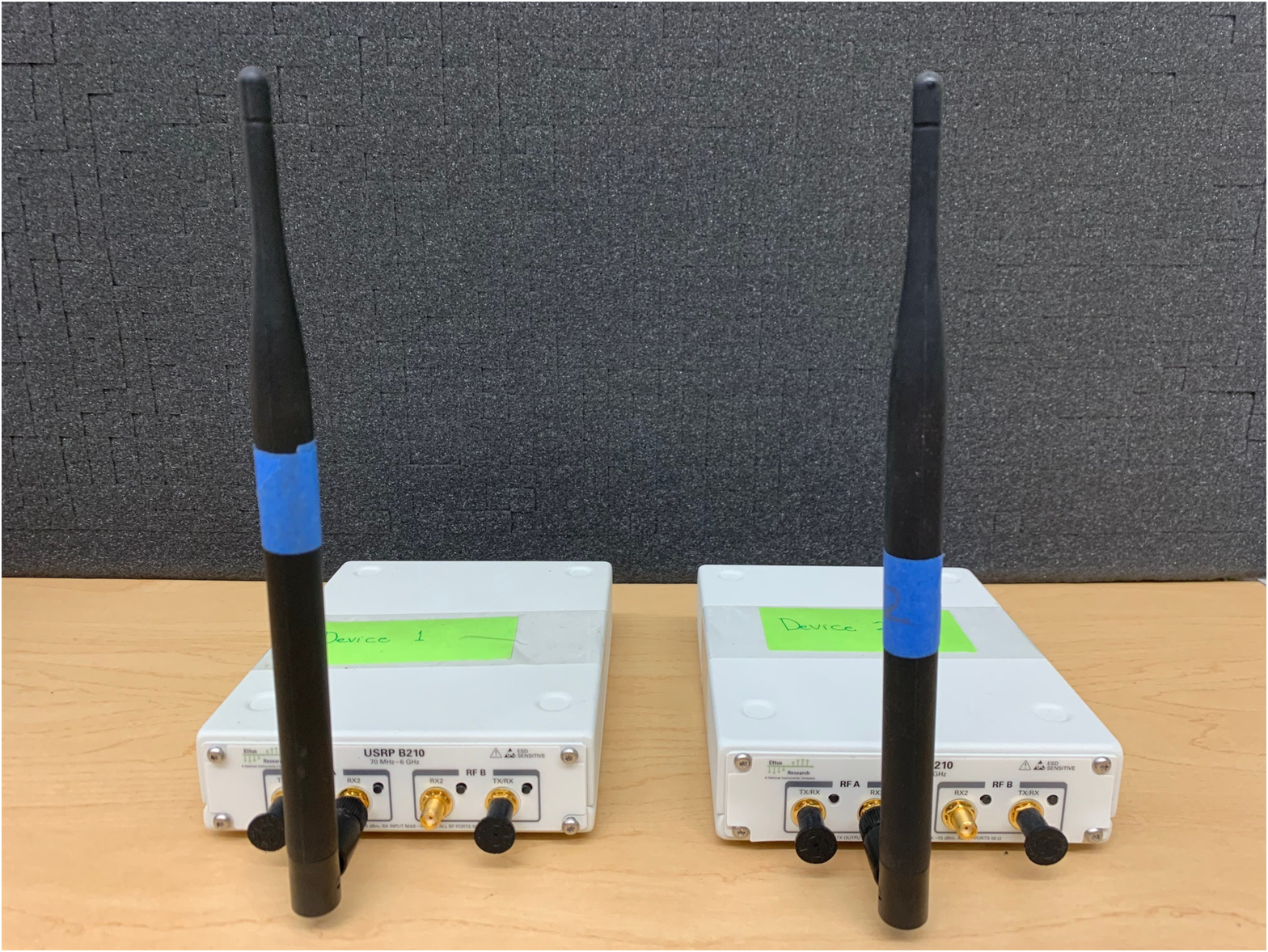}\label{subfig:Receivers}}\\
   &
\subfloat[PyCom LoRa transmitters]{
   \includegraphics[width=0.5\columnwidth]{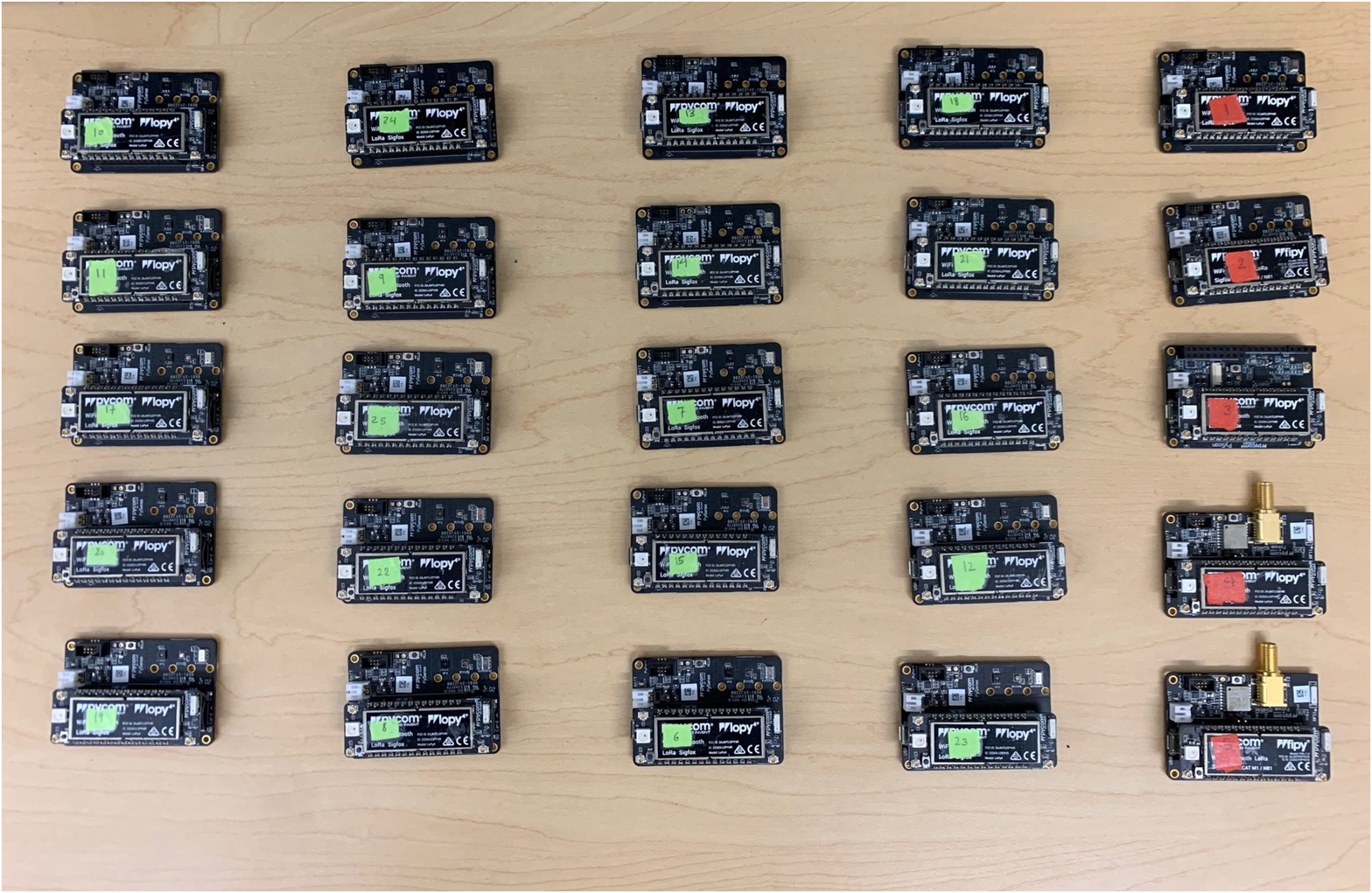}\label{subfig:Transmitters}}\\
\end{tabular}
\caption{(a) Twin network, (b) and (c) testbed hardware.}
\label{fig:network}
\end{figure}

This section describes the testbed used for collecting the RF datasets and the experimental setups used for evaluating \proposed\ under three different portability scenarios.

\subsection{Testbed}
In order to evaluate \proposed, real RF data was collected using a testbed of devices. This testbed contains 25 almost identical PyCom IoT devices used as transmitters: 23 Lopy4 boards and 2 Fipy boards on top of 22 Pysense sensor shields, 2 Pytrack sensor shields, and 1 Pyscan sensor shield (pictured in Fig.~\ref{subfig:Transmitters}). The testbed also uses 2 USRP B210 SDRs (Software Defined Radios) as receivers to collect data (pictured in Fig.~\ref{subfig:Receivers}). 

All of the datasets collected using this testbed (outlined in the next section) use the following settings and processes unless otherwise stated: the PyCom transmitters were configured to transmit using the LoRa protocol, with a center frequency of 915 MHz, spreading factor of 7, and bandwidth of 125KHz. The USRP B210 receivers were configured to sample at a center frequency of 915 MHz, at a rate of 1M samples per second. Each transmission sends the same message, and lasts for 20s. This produces 20M complex-valued samples for each transmission received. This data was stored in raw-IQ format, using GNURadio to process the data.

Regardless of the environment in which data was collected, the transmitters and receivers were always placed about 5m apart (with the exception of `Wired' data, which is collected by connecting transmitters to receivers by physical wire). All data collected using this testbed will be made publicly available.

\subsection{Experimental Scenarios}

\begin{table}
\centering
\begin{tabular}{|c|c|c|c|c|c|c|}
    \hline
    \textbf{Portability} & \textbf{Env.} & \begin{tabular}{@{}c@{}}\textbf{Multi-RX} \\ \textbf{Collection}\end{tabular} & \begin{tabular}{@{}c@{}}\textbf{\#} \\ \textbf{TX}\end{tabular} & \begin{tabular}{@{}c@{}}\textbf{\#} \\ \textbf{RX}\end{tabular} & \begin{tabular}{@{}c@{}}\textbf{\#} \\ \textbf{Days}\end{tabular} & \begin{tabular}{@{}c@{}}\textbf{\# TX} \\ \textbf{Config.}\end{tabular}\\
    \hline
    Hardware & Indoor & Same Tx & 25 & 2 & 1 & 1 \\
    Hardware & Outdoor & Same Tx & 25 & 2 & 1 & 1 \\
     Hardware & Wired & Same Tx & 25 & 2 & 1 & 1 \\
    Hardware & Indoor & Diff. Tx & 25 & 2 & 1 & 1 \\
    Hardware & Outdoor & Diff. Tx & 25 & 2 & 1 & 1 \\
    Hardware & Wired & Diff. Tx & 25 & 2 & 1 & 1 \\
    \hline
    Channel & Outdoor & N/A & 25 & 1 & 5 & 1 \\
    \hline
    Config. & Indoor & N/A & 25 & 1 & 1 & 4 \\
    \hline  
\end{tabular}
\caption{Summary of the datasets used for evaluation.}
\label{table:scenarios}
\end{table}

The collected LoRa RF datasets and the studied portability scenarios are summarized in Table~\ref{table:scenarios} and described next.

{\bf Hardware portability dataset scenario.} Six different datasets were collected to evaluate the hardware portability scenario. All of these datasets involve the same 2 receivers and 25 transmitters, and each contains 2 recorded transmissions from each transmitter: one at each receiver. 

\begin{figure}
\begin{tabular}{c|c}
\centering
\subfloat[Same Transmission]{
   \includegraphics[width=0.45\columnwidth]{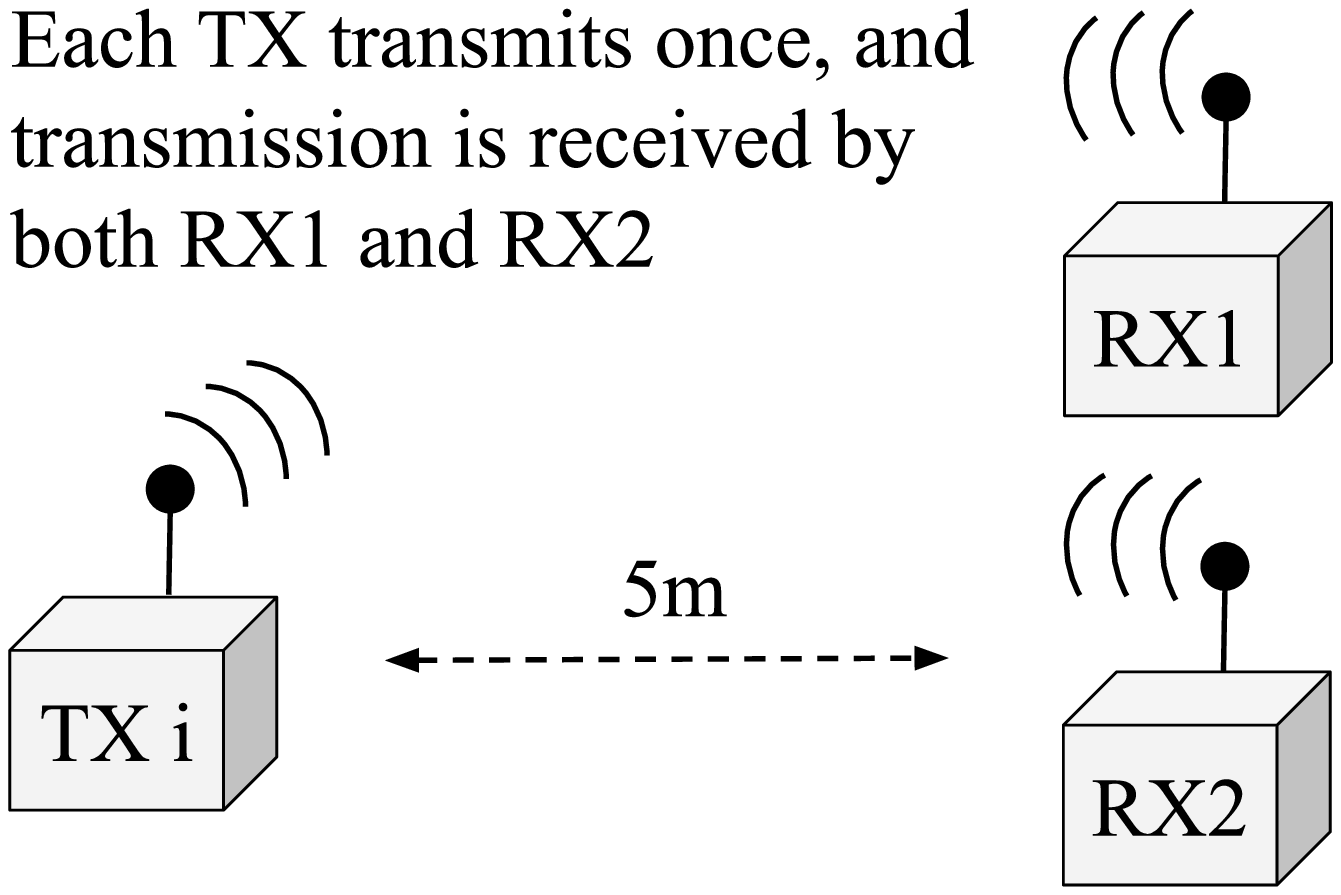}\label{subfig:sameTX}}     & 
\subfloat[Different Transmission]{
   \includegraphics[width=0.4\columnwidth]{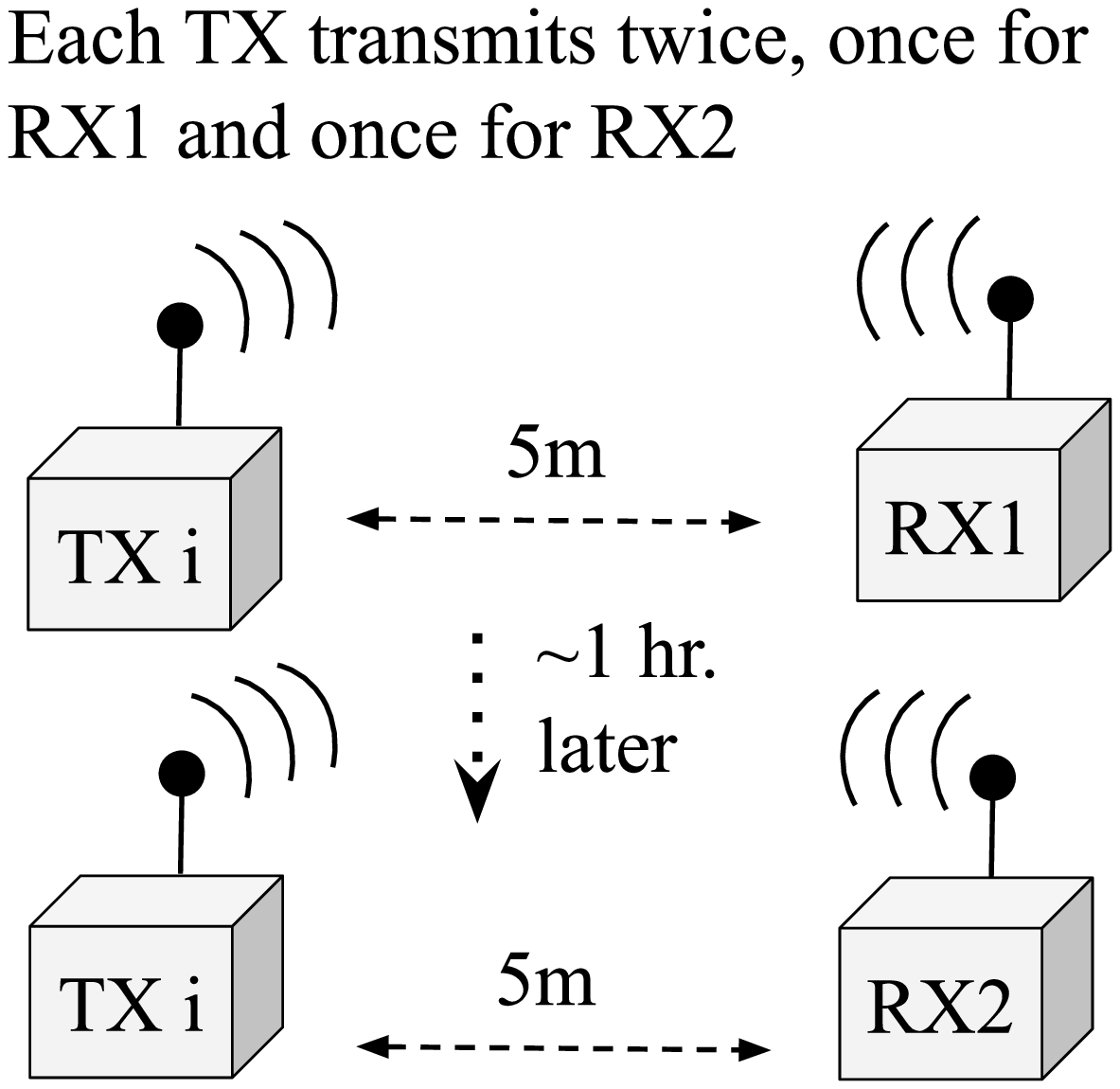}\label{subfig:diffTX}}
\end{tabular}
\caption{Methods of collecting RF data with two receivers. }
\label{fig:same_trans_diff_trans}
\end{figure}

These six datasets differ in terms of the environment they were collected in as well as the `Multi-RX Collection' method used. Datasets where the `Multi-RX Collection' method is `Diff. Tx' were collected by capturing two \underline{different} transmissions from each device, one at each receiver as shown in Fig.~\ref{subfig:diffTX}. Datasets where this method is `Same Tx' were collected by capturing the \underline{same} single transmission from each device at both receivers as shown in Fig.~\ref{subfig:sameTX}.

{\bf Channel portability dataset scenario.} This dataset includes 5 transmissions from each of the 25 transmitters, with each of the 5 transmissions collected on a different day. Our intent is to have this dataset represent transmissions under 5 different wireless channel conditions.
 
{\bf Protocol configuration portability dataset scenario.} This dataset includes four transmissions from each of 25 transmitters, each of the four transmissions using a different configuration. The configurations amount to using different LoRa spreading factors and are detailed in Table~\ref{table:lora}. Note that all other datasets use only Config 1.

\begin{table}
\centering
\small
\begin{tabular}{|c|c|c|c|c|c|}
    \hline
    \textbf{Config.} & \begin{tabular}{@{}c@{}}\textbf{Spreading} \\ \textbf{Factor}\end{tabular} & \textbf{BW} & \begin{tabular}{@{}c@{}}\textbf{Tx} \\ \textbf{Power}\end{tabular} &  \begin{tabular}{@{}c@{}}\textbf{Coding} \\ \textbf{Rate}\end{tabular}&\textbf{Bit Rate}\\
    \hline
    1 & 7 & 125kHz & 20dBm & 4/5 & 5470bps \\
    2 & 8 & 125kHz & 20dBm & 4/5 & 3125bps \\
    3 & 11 & 125kHz & 20dBm & 4/5 & 537bps \\
    4 & 12 & 125kHz & 20dBm & 4/5 & 293bps \\
    \hline  
\end{tabular}
\caption{Different LoRa transmitter configurations.}
\label{table:lora}
\end{table}

\section{Experimental Results}
\label{sec:performance}
We first describe the specific implementation of the twin network as well as the metrics to be used to evaluate \proposed. Then, we explain the results obtained by testing \proposed\ under the portability scenarios outlined in the previous section.

\subsection{Neural Network Implementation}
The network used for this work is inspired by one of the straightforward CNNs described in~\cite{alshawabkainfocom} and is depicted in Fig.~\ref{subfig:twinNet}. Note that this represents one of the two identical halves of the twin network. The network was implemented using the Python `PyTorch' library and uses a total of four 1-D convolutional layers, making use of 1-D max-pooling and batch normalization layers after each pair of convolutional layers. It also uses two fully-connected layers at its output, ending with 12 neurons. This produces a 12-dimensional output point for each input example. Leaky ReLU is used as the activation function throughout the network. 

The inputs to the network are presented in the form of complex IQ (In-phase, Quadrature) samples,  a common format for representing RF signals. For this work, a sequence length of 128 complex samples was considered as one input example. Additionally, two different input channels were considered, one for I samples and one for Q samples. Thus, the final input shape for the network is 2x128. 

The triplet loss function is used to train the network with a margin value of 0.1, and the ``difficult triplet selection" strategy described earlier is employed to select triplets for training. The optimizer chosen was Stochastic Gradient Descent (SGD) with Momentum set to 0.9. Batch size was set to 64, and the learning rate for each model was tuned over the range 1e-1 to 1e-6. The networks were allowed to train for 100 epochs and the model was saved at the best performing epoch.

For baseline comparison, a `vanilla' network was defined using half of a twin network (Fig.~\ref{subfig:twinNet}) and having 10 neurons in the final layer (where 10 is the number of transmitters used during testing). The same hyper-parameters described for the proposed network are used for training, but the `vanilla' network uses cross-entropy loss instead of triplet-loss.

\subsection{Performance Evaluation Metrics}
To assess \proposed's performance, three metrics are used: (i) the averaged Area Under the Receiver Operating Characteristic ({\bf Avg. AUROC}); (ii) the averaged True Positive Rate ({\bf Avg. TPR}); and (iii) the averaged False Positive Rate ({\bf Avg. FPR}). These three metrics are further explained next.

\begin{figure}
    \centering
    \hfill
    \subfloat[ROC curve.]{
    \includegraphics[width=0.25\columnwidth,height =.25\columnwidth]{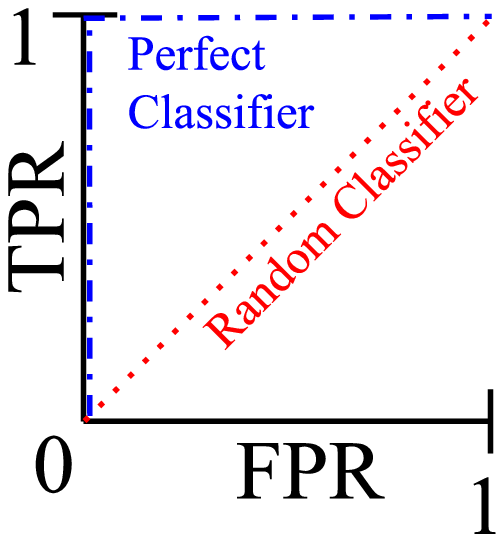}\label{subfig:AUROC_example}}
    \hfill
    \subfloat[Possible decision outcomes.]{
    \includegraphics[width=0.6\columnwidth,height =.25\columnwidth]{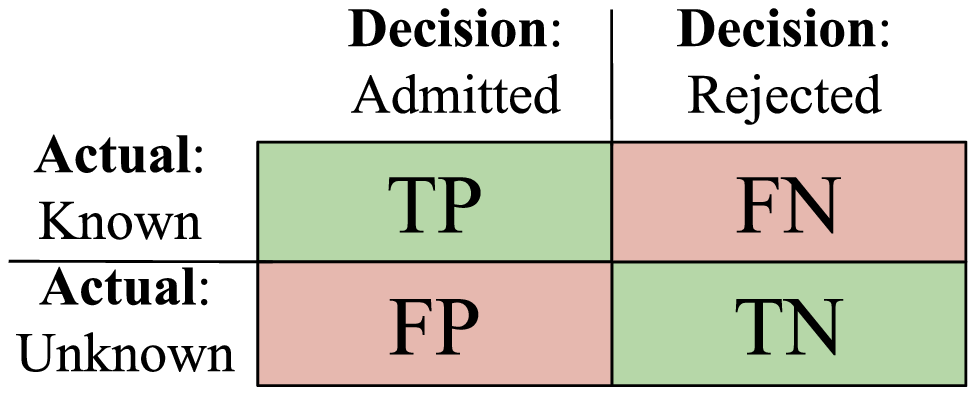}\label{subfig:TPR_FPR_example}}
    \caption{\small Visualization of performance metrics.}
    \label{fig:metrics}
\end{figure}

 The TPR and FPR metrics are calculated using the exact decision making process defined in Algorithm~\ref{alg:decisionOpen} as, $TPR = {TP}/{(TP + FN)}$ and $FPR = {FP}/{(FP + TN)}$, where:
 \begin{itemize}
\item True Positive (TP) is the number of tested known devices that the model predicts as known (correctly)
\item False Negative (FN) is the number of tested known devices that the model predicts as unknown (incorrectly)
\item False Positive (FP) is the number of tested unknown devices that the model predicts as known (incorrectly)
\item True Negative (TN) is the number of tested unknown devices that the model predicts as unknown (correctly)
\end{itemize}
These quantities are also shown visually in Fig.~\ref{subfig:TPR_FPR_example}.

The AUROC metric allows for measuring the performance of a binary classifier (i.e., classifying an input/device as known or unknown) without having to specify a particular threshold. Instead, decision scores are collected and the TPR and FPR are calculated for all possible thresholds on these scores. The AUROC is then defined as the area under the ROC curve resulting from plotting FPR vs. TPR for all thresholds. Fig.~\ref{subfig:AUROC_example} depicts an example of this curve. 
Note that the open-set evaluations using AUROC do not strictly follow Algorithm~\ref{alg:decisionOpen} defined earlier, since this algorithm uses specific thresholds (the $\bm{Distances}$). Instead, for models using \proposed, the AUROC decision scores correspond to the minimum distance between $\bm{InputPoint}$ and a $\bm{Centroid}$, and for the `vanilla' models, the decision scores correspond to the maximum logit value for each example, which is commonly used in open set classification.

The reported evaluation metrics (AUROC, TPR and FPR) are averaged over 5 trials since performance can depend on which unknown devices are used for testing. For each trial, an equal number of examples were drawn from 5 random known devices and 5 random unknown devices. 

\subsection{Evaluation Results}
\proposed\ is evaluated in an open-set device authentication setting, which requires it to decide if inputs originated from a known (authorized) or unknown (unauthorized) device. 
All models evaluated in this section were trained and tested using data transmissions collected from the same 10 wireless IoT transmitters, with 75\% of the data used for training and 25\% used for testing (each transmission contains 156,250 examples of size 128x2). Test data from the other 15 IoT transmitters was used to represent the unknown/unauthorised devices. In all tests, the amount of calibration data used ($N$ in Algorithm~\ref{alg:calibration}) was set to 10\% of the size of the training data used to initially train the model (11,719 examples), and the number of input examples used to form the `input point' ($M$ in Algorithm~\ref{alg:decisionOpen}) was set to 10. It is also worth noting here that on all AUROC results presented in this section, a dashed line is plotted at 0.5 to indicate the performance of a random classifier/authentication.

Additionally, for comparison to \proposed, `vanilla' models are also evaluated in this section. Note that the `vanilla' models were only evaluated using the AUROC metric.

\subsubsection{Hardware Portability}

\begin{figure*}[t]
\centering
\begin{tabular}{m{0.65\columnwidth} m{0.25\columnwidth} m{0.65\columnwidth} m{0.25\columnwidth}}
\multirow{2}{*}[1em]{
\renewcommand{\thesubfigure}{a}
\subfloat[AUROC - Different Tx]{
   \includegraphics[width=0.65\columnwidth,valign=t]{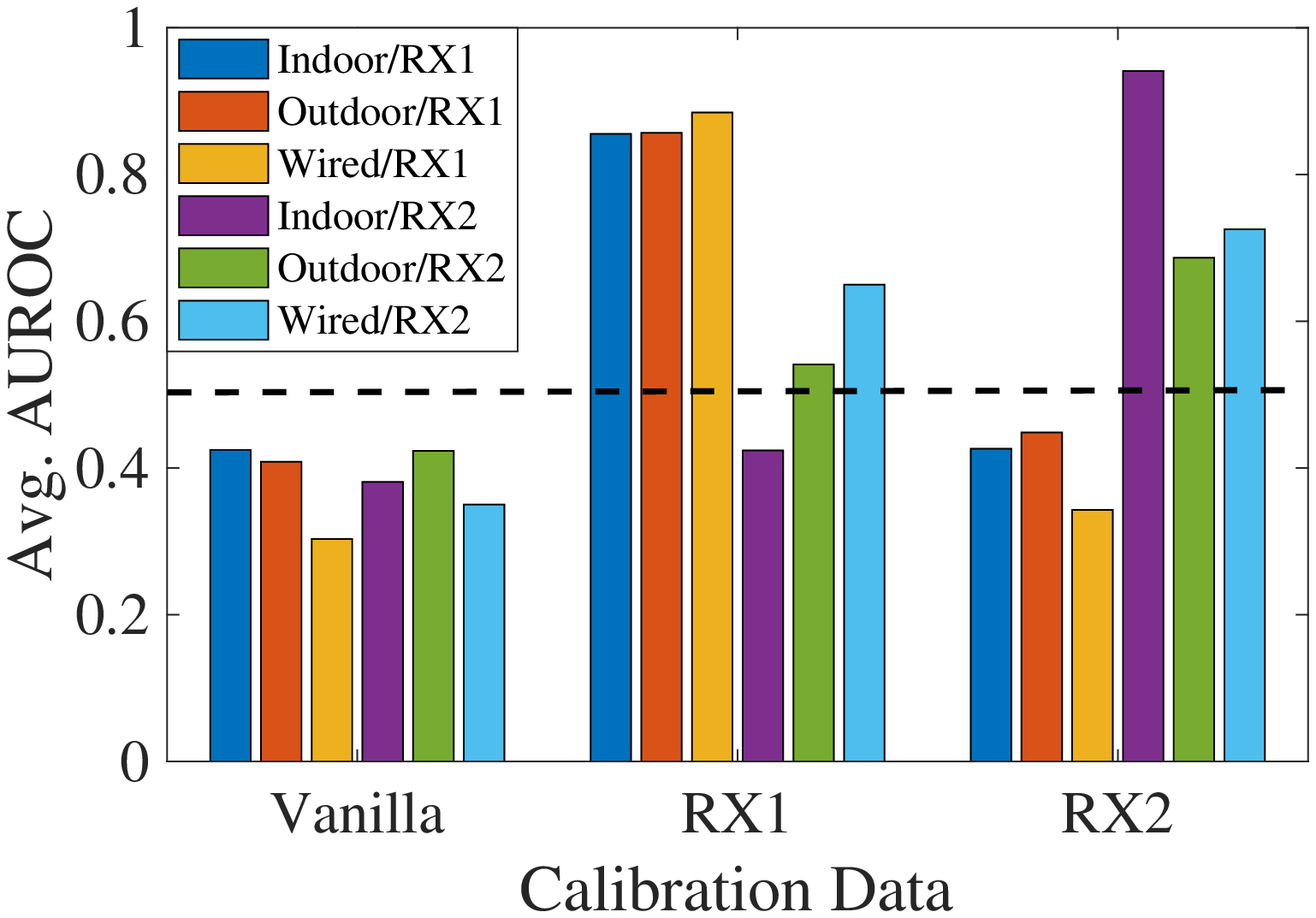}\label{subfig:rx1-difftx-AUROC}}
   }
   &
    \renewcommand{\thesubfigure}{b}
\subfloat[TPR - Diff. Tx]{
   \includegraphics[width=0.3\columnwidth]{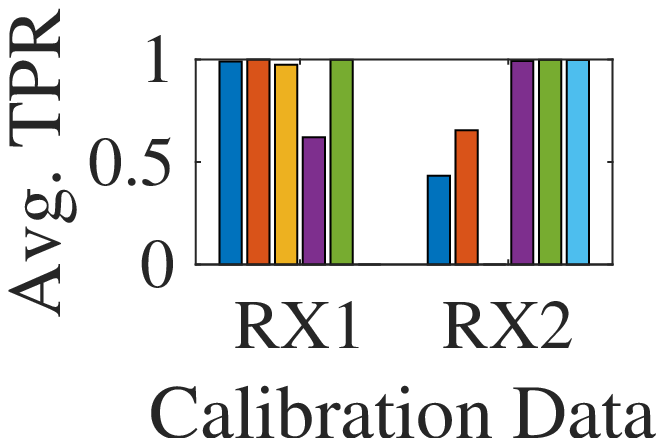}\label{subfig:rx1-difftx-TPR}}
   &
   \multirow{2}{*}[1em]{
   \renewcommand{\thesubfigure}{d}
\subfloat[AUROC - Same Tx]{
   \includegraphics[width=0.65\columnwidth,valign=t]{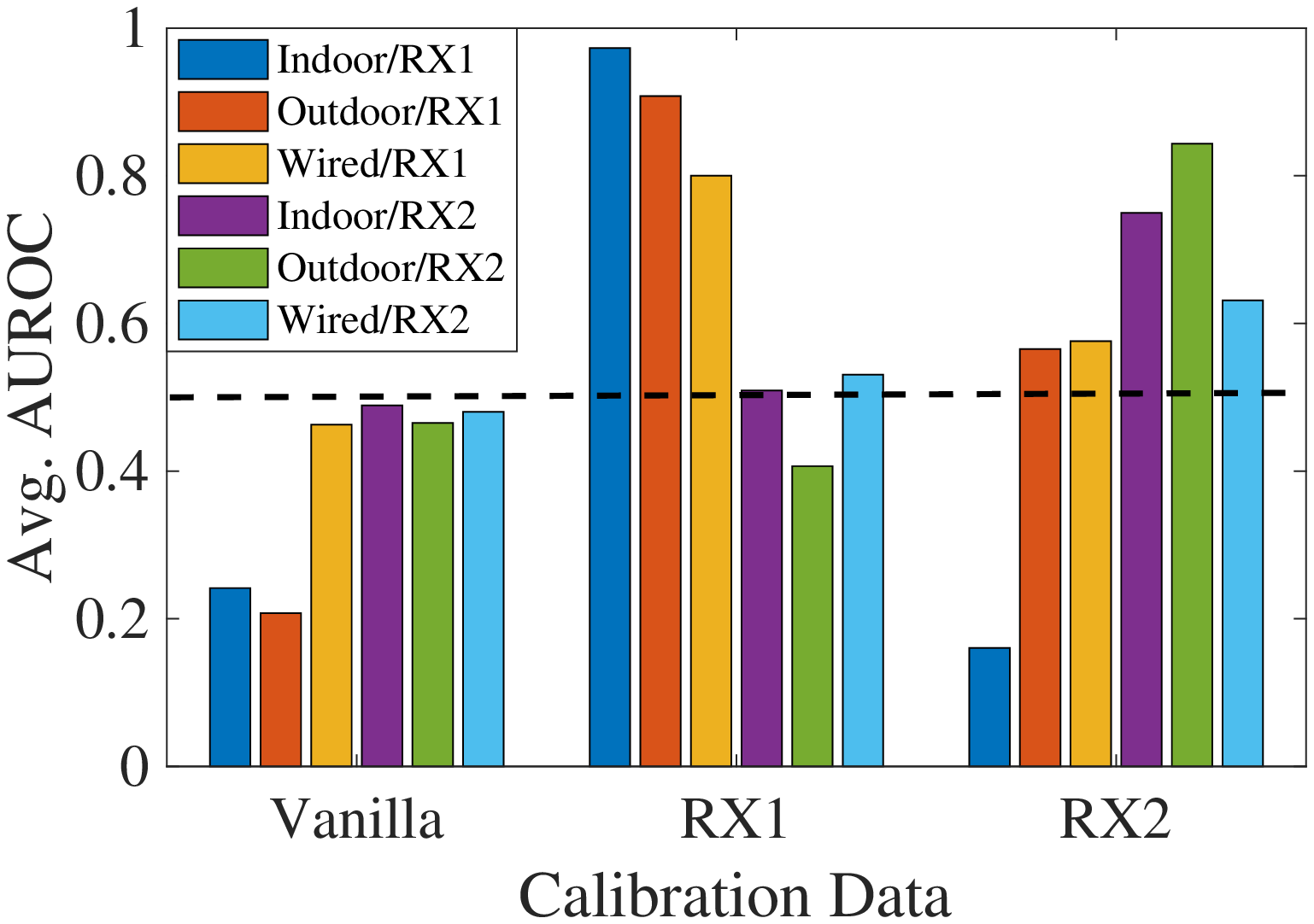}\label{subfig:rx1-sametx-AUROC}}
   }
   &
   \renewcommand{\thesubfigure}{e}
\subfloat[TPR - Same Tx]{
   \includegraphics[width=0.3\columnwidth]{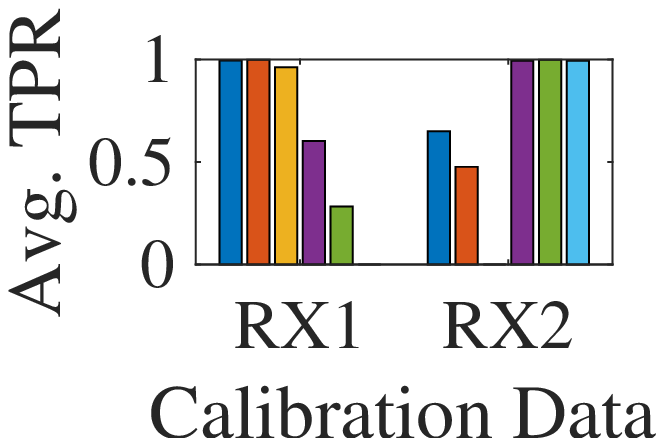}\label{subfig:rx1-sametx-TPR}}
\\
   &
   \renewcommand{\thesubfigure}{c}
\subfloat[FPR - Diff. Tx]{
   \includegraphics[width=0.3\columnwidth]{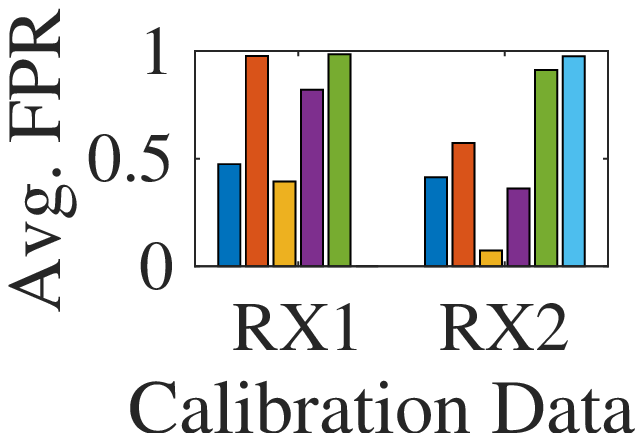}\label{subfig:rx1-difftx-FPR}}
   &
   &
   \renewcommand{\thesubfigure}{f}
\subfloat[FPR - Same Tx]{
   \includegraphics[width=0.3\columnwidth]{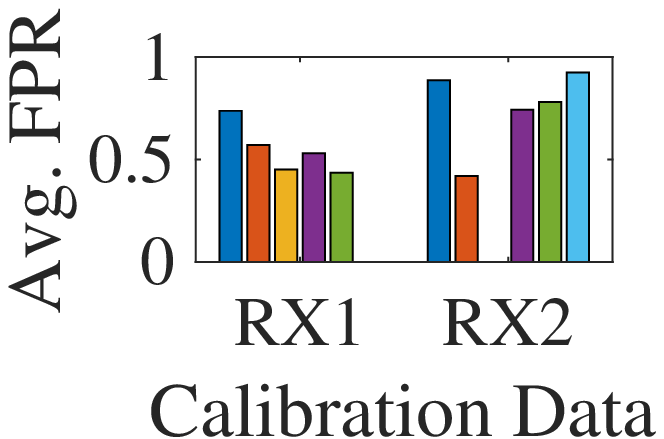}\label{subfig:rx1-sametx-FPR}}
   \\
   
   \comment{
   \\
\renewcommand{\thesubfigure}{g}
\subfloat[AUROC - Train RX2 - Diff. Tx]{
   \includegraphics[width=0.98\columnwidth]{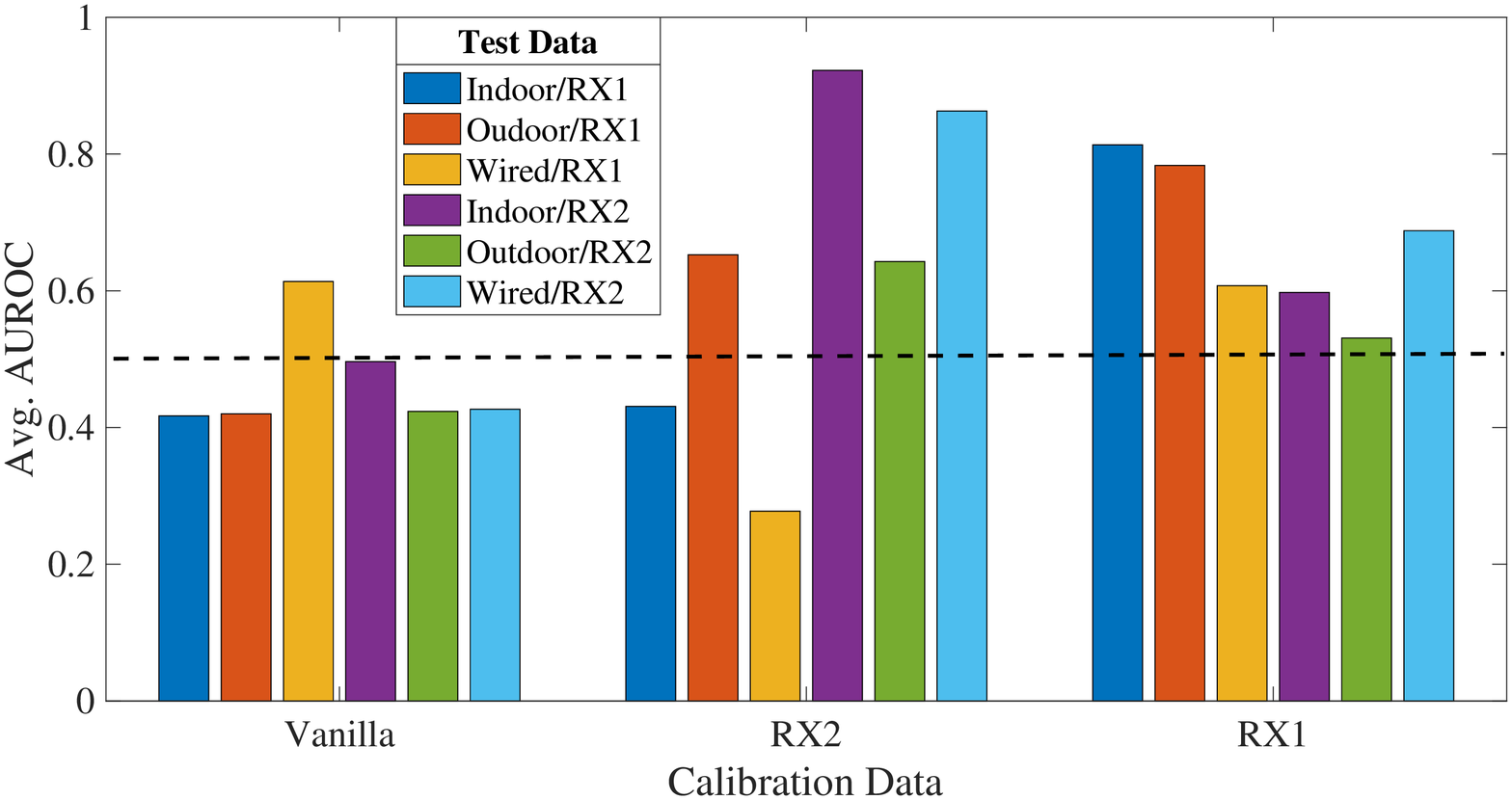}\label{subfig:rx2-difftx-AUROC}}
\renewcommand{\thesubfigure}{j}
\subfloat[AUROC - Train RX2 - Same Tx]{
   \includegraphics[width=0.98\columnwidth]{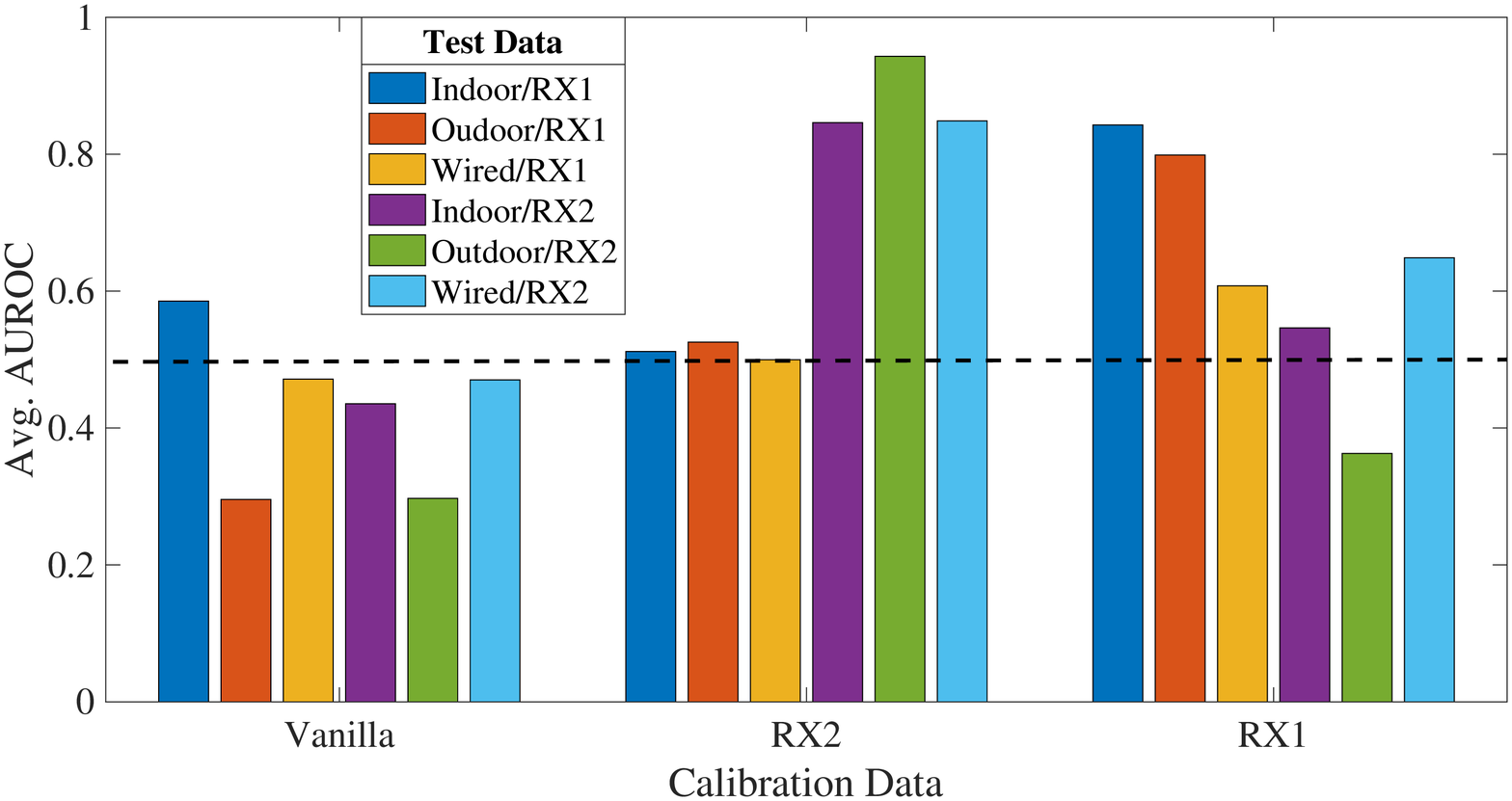}\label{subfig:rx2-sametx-AUROC}}\\
\renewcommand{\thesubfigure}{h}
\subfloat[TPR - Train RX2 - Diff. Tx]{
   \includegraphics[width=0.48\columnwidth]{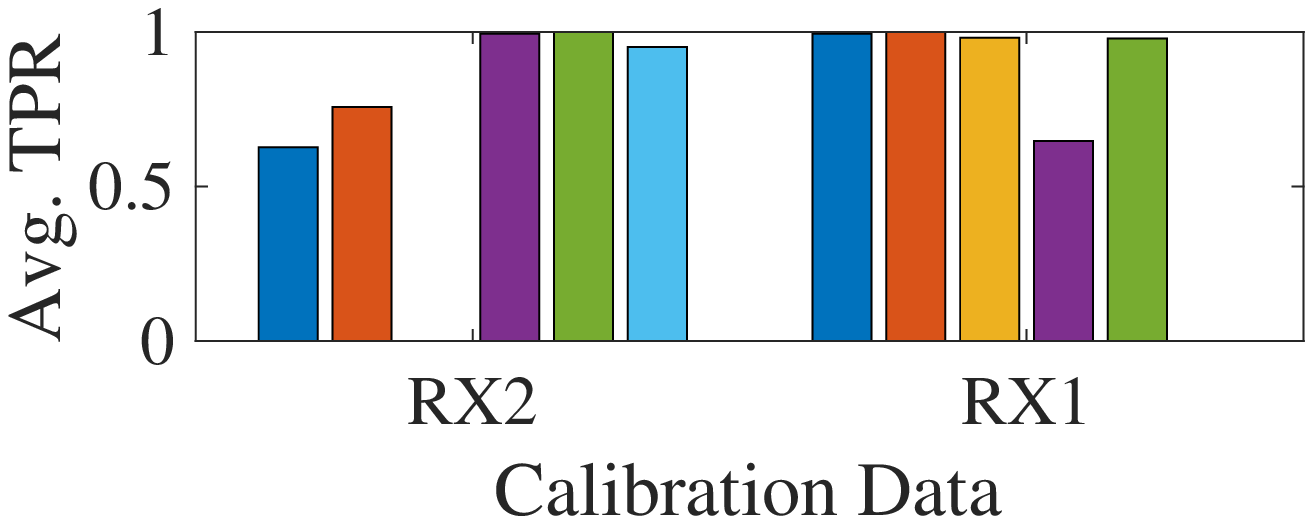}\label{subfig:rx2-difftx-TPR}}
\renewcommand{\thesubfigure}{i}
\subfloat[FPR - Train RX2 - Diff. Tx]{
   \includegraphics[width=0.48\columnwidth]{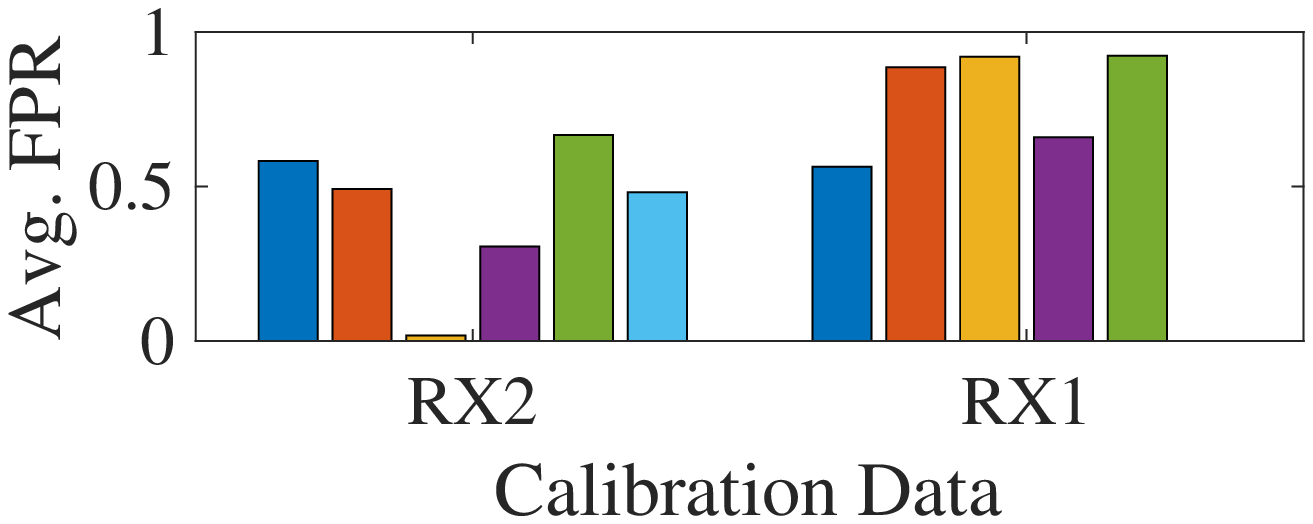}\label{subfig:rx2-difftx-FPR}}
\renewcommand{\thesubfigure}{k}
\subfloat[TPR - Train RX2 - Same Tx]{
   \includegraphics[width=0.48\columnwidth]{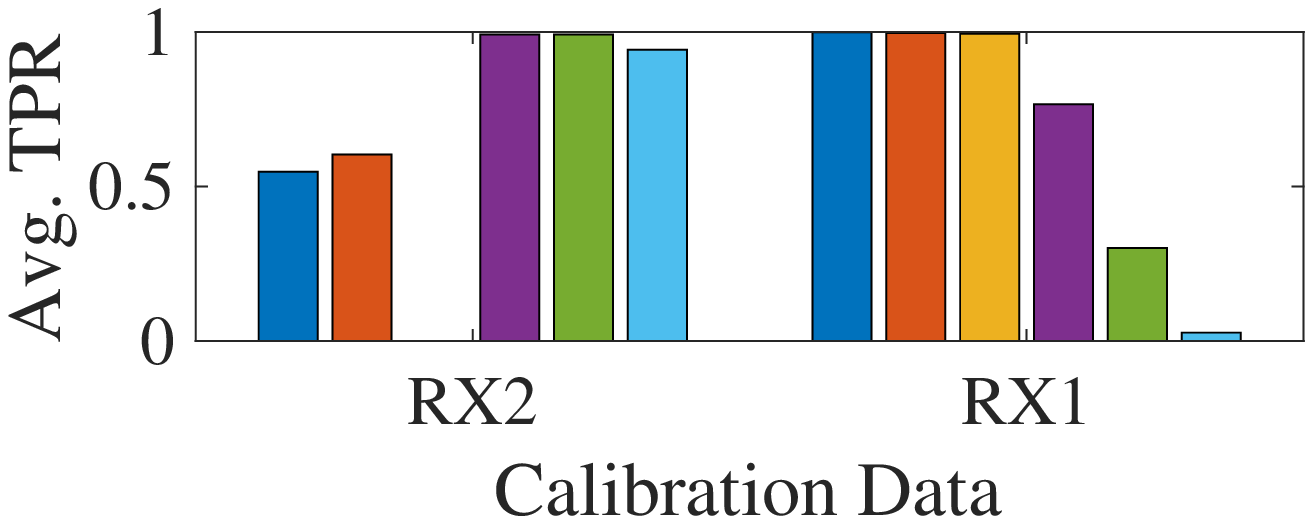}\label{subfig:rx2-sametx-TPR}}
\renewcommand{\thesubfigure}{l}
\subfloat[FPR - Train RX2 - Same Tx]{
   \includegraphics[width=0.48\columnwidth]{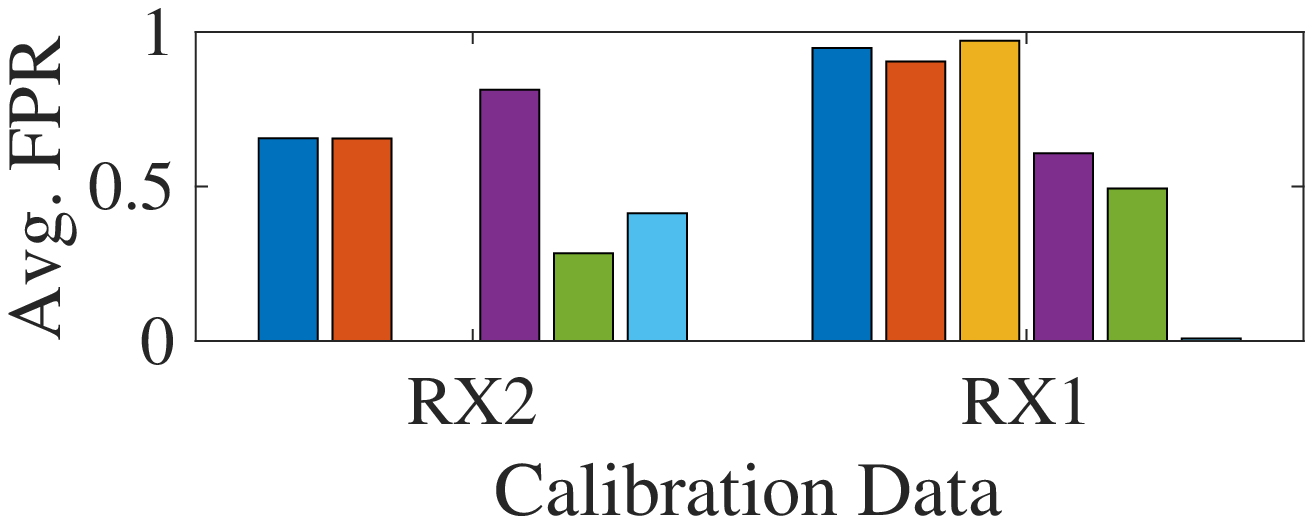}\label{subfig:rx2-sametx-FPR}}
   }
\end{tabular}
\caption{\textbf{Hardware Portability}: Models trained on data collected at RX1. $N$ = 10\% and $M$ = 10.}  
\label{fig:rx_open}
\end{figure*}

We begin by assessing \proposed's ability in achieving hardware portability. For this, for each of the six studied hardware portability datasets listed in Table~\ref{table:scenarios}, \proposed\ was trained on data collected at RX1. Then, \proposed\ is tested with data collected at both receivers, RX1 and RX2, while being: (i) calibrated with data collected at RX1, and (ii) calibrated with data collected at RX2 (receiver different from that used for collecting training data).
 
Figs.~\ref{subfig:rx1-difftx-AUROC}-\ref{subfig:rx1-difftx-FPR} show the results of these tests for the `Diff. Tx' method, where data captured by RX1 and RX2 for each transmitter is done during two separate transmissions. Figs.~\ref{subfig:rx1-sametx-AUROC}-\ref{subfig:rx1-sametx-FPR} show the same results but for the `Same Tx' method, where data is collected at RX1 and RX2 by capturing the same transmission from each device. 
Note: models trained on RX2 data as opposed to RX1 data were also evaluated and showed similar results, and are omitted due to space limitations.

The main trend present for the AUROC results is that in most cases, better AUROC is achieved when testing on the same receiver for which the model is calibrated. This can be seen in Fig.~\ref{subfig:rx1-difftx-AUROC}, for instance, where the results when calibrating/testing on RX1 data and the results when calibrating/testing on RX2 data show higher AUROC when compared to the other tests. It is also notable that the `vanilla' models perform far worse than \proposed\ in terms of AUROC when the models that use \proposed\ are calibrated and tested on the same receiver. This is indicative of the potential for \proposed\ to achieve hardware portability for device authentication.

The TPR results shown in Figs.~\ref{subfig:rx1-difftx-TPR} and~\ref{subfig:rx1-sametx-TPR} provide even stronger evidence of the main trend that existed for the AUROC metric, where performance is high when the model is calibrated/tested on the same receiver. These results indicate that \proposed\ always achieves a high TPR, meaning it almost always correctly `admits' examples from known devices.

In the FPR results, the trend is less clear. Ideally, models calibrated for a particular receiver would achieve a low FPR when tested on data from that receiver, but this is not always the case. For example, in Figs.~\ref{subfig:rx1-difftx-FPR} and~\ref{subfig:rx1-sametx-FPR}, it can be seen that there are several cases where a model is calibrated for data from a receiver, but still achieves a higher FPR when tested on data from that receiver than when it is tested with data from a receiver for which it is not calibrated. For a particular example, notice the last two bars of the tests for models calibrated with RX2 data in Fig.~\ref{subfig:rx1-difftx-FPR}. These are both tests on RX2 data, so good performance would be expected, and thus low FPR. Instead these tests have the highest FPR of any test for models calibrated to RX2 in the figure.

This seemingly counter-intuitive result can be explained by considering the meaning of FPR, and what exactly the model is attempting to do in the latent space during evaluation. Recall that when a model is calibrated using a particular set of data it forms a $\bm{Centroid}$ for each device in its latent output space. To achieve low FPR (which is desirable), the examples from unknown devices must form distinct `clusters' in the latent space far enough away from the $\bm{Centroids}$ so that the model can distinguish known from unknown. 

In the case where the model is calibrated/tested on data from the same domain, the `clusters' from both the known and unknown devices will likely be closer to the $\bm{Centroids}$ in the latent space than the case where the model is calibrated/tested with data from two different domains. This means that when testing on a different domain than calibration, the model will more easily be able to `reject' examples from unknown devices (achieving a lower FPR), but will also have a more difficult time `admitting' examples from known devices (also achieving a lower TPR). Thus, achieving a low FPR is generally more difficult when testing on the same domain as calibration. 

It is also worth noting that even when the FPR produced by calibrating and testing on data from the same receiver is not the highest FPR, the result is still not as low as would be desired. For example, observe Fig.~\ref{subfig:rx1-difftx-FPR}. Notice that the FPR for the model calibrated and tested on RX2 data from an `Indoor' environment is among the lowest for all models calibrated with RX2 data, but that it is still in excess of 0.25. This is indicative of an issue with the $\bm{Distances}$ chosen during calibration and used for open-set decision making. It indicates that these $\bm{Distances}$ may be too large, producing a much higher TPR as well as a relatively high FPR. 

\begin{figure}
\centering
\begin{tabular}{m{0.55\columnwidth} m{0.4\columnwidth}}
\multirow{2}{*}[1em]{
\subfloat[AUROC]{
   \includegraphics[width=0.60\columnwidth]{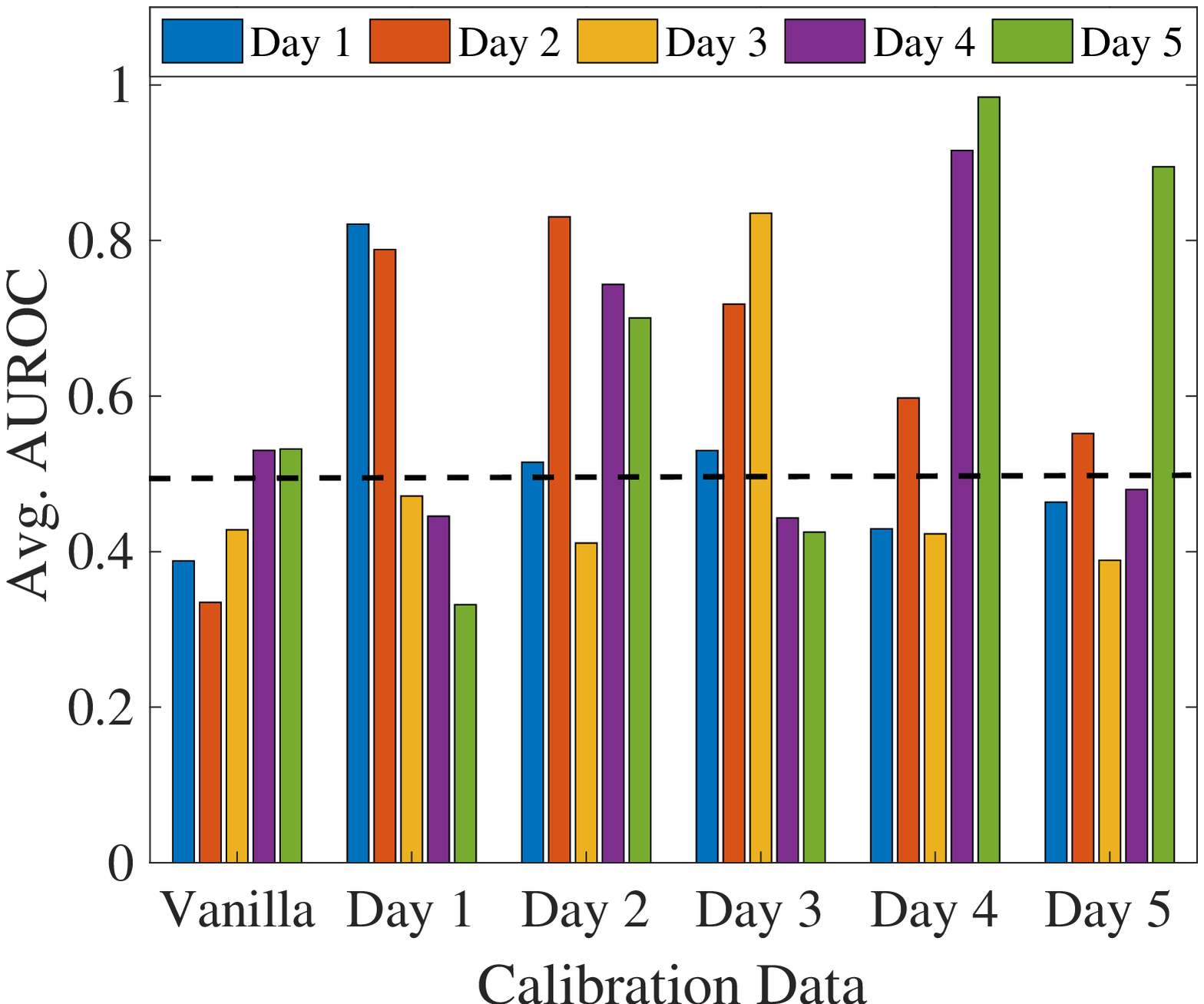}\label{subfig:day1_AUROC}}
   }
   &
   \subfloat[TPR]{
   \includegraphics[width=0.4\columnwidth]{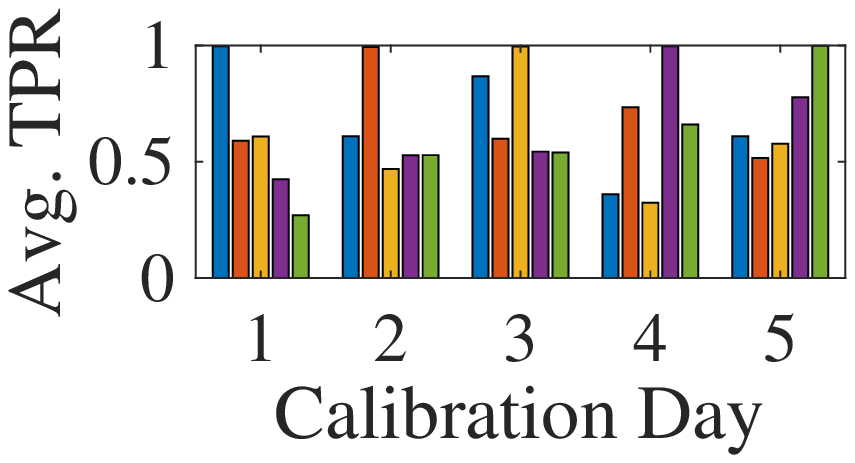}\label{subfig:day1_TPR}}
   \\
    &
\subfloat[FPR]{
   \includegraphics[width=0.4\columnwidth]{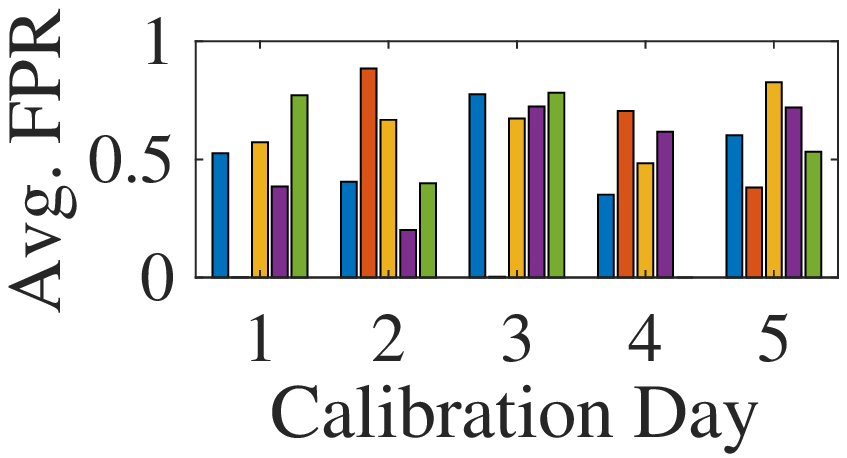}\label{subfig:day1_FPR}}\\
   
   \comment{
   \\
\subfloat[AUROC - Train Day 2]{
   \includegraphics[width=0.98\columnwidth]{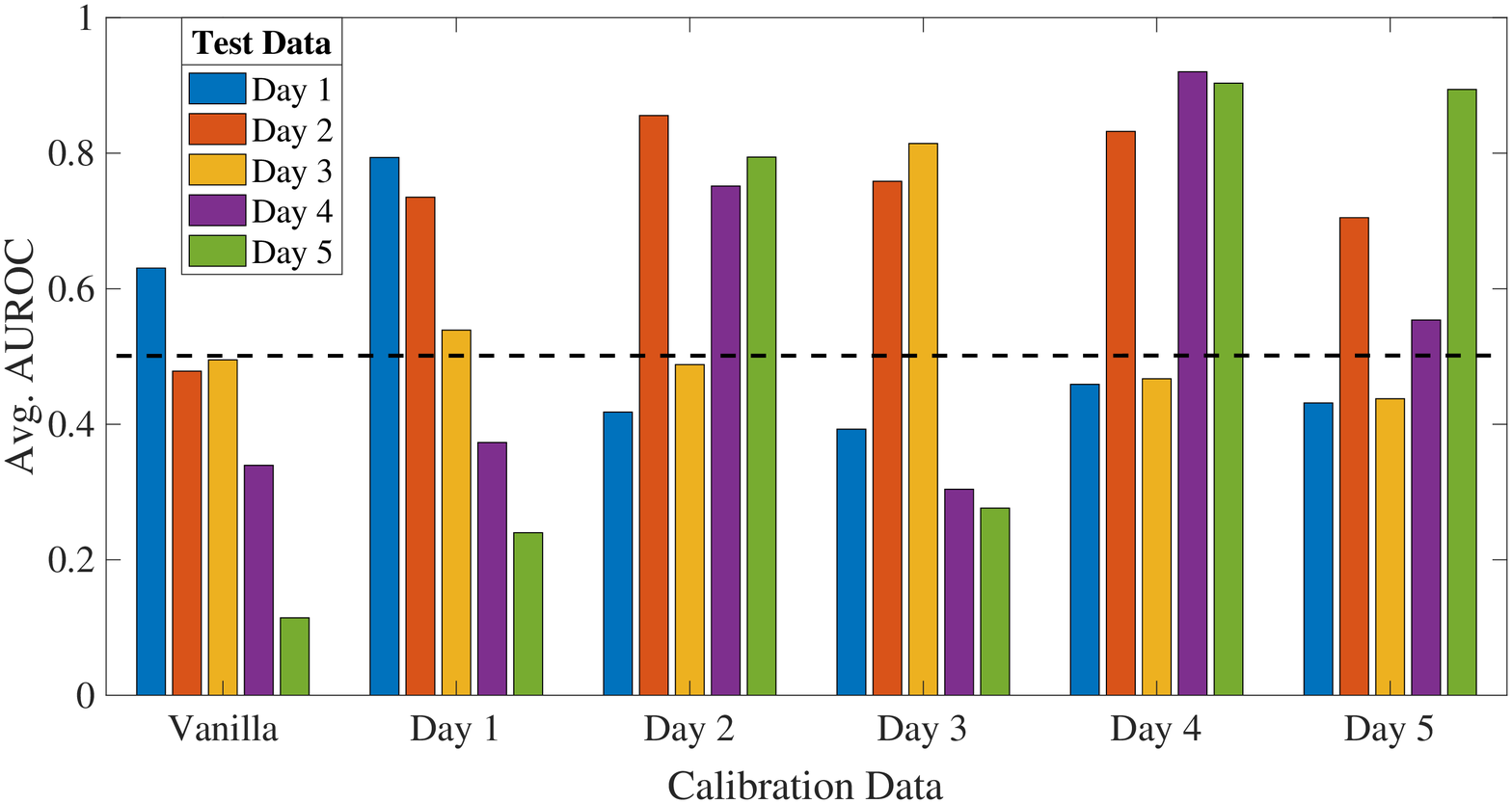}\label{subfig:day2_AUROC}}\\
\subfloat[TPR - Train Day 2]{
   \includegraphics[width=0.48\columnwidth]{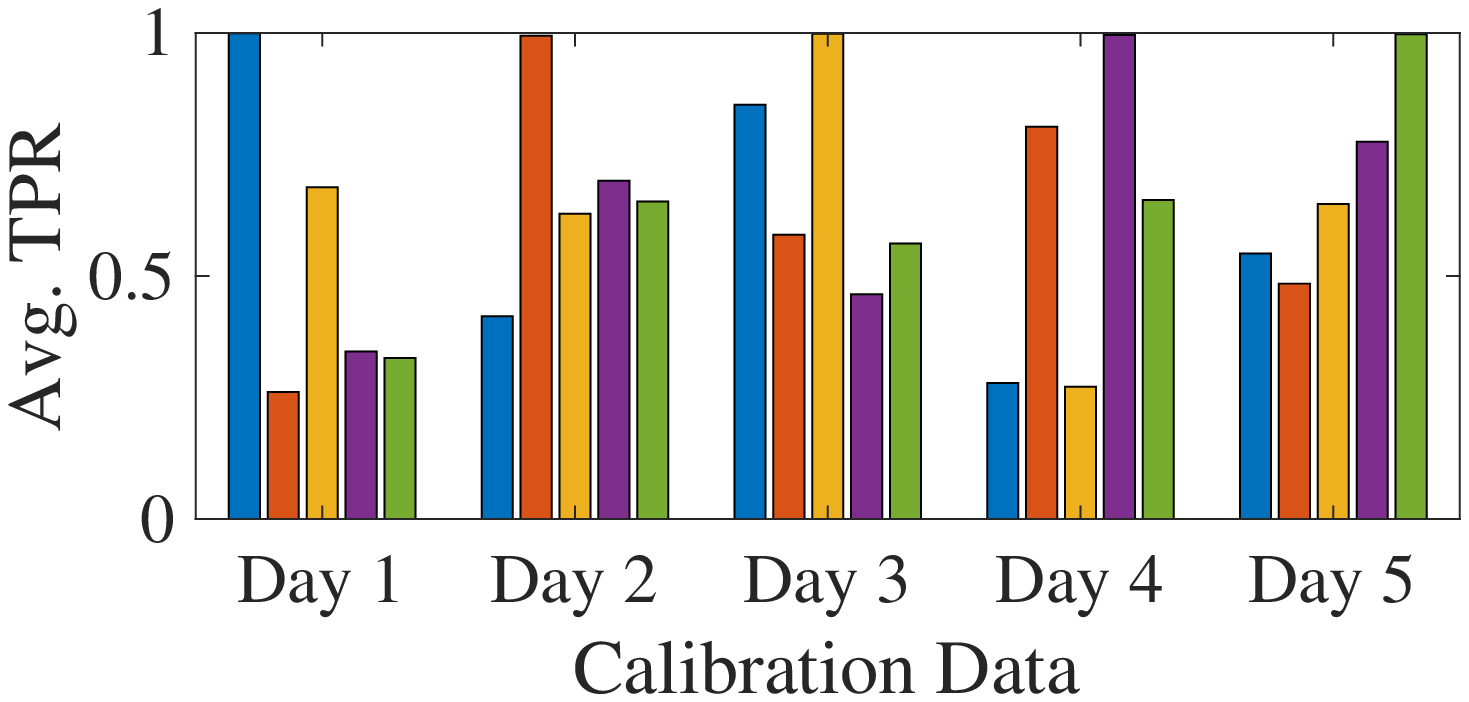}\label{subfig:day2_TPR}}
\subfloat[FPR - Train Day 2]{
   \includegraphics[width=0.48\columnwidth]{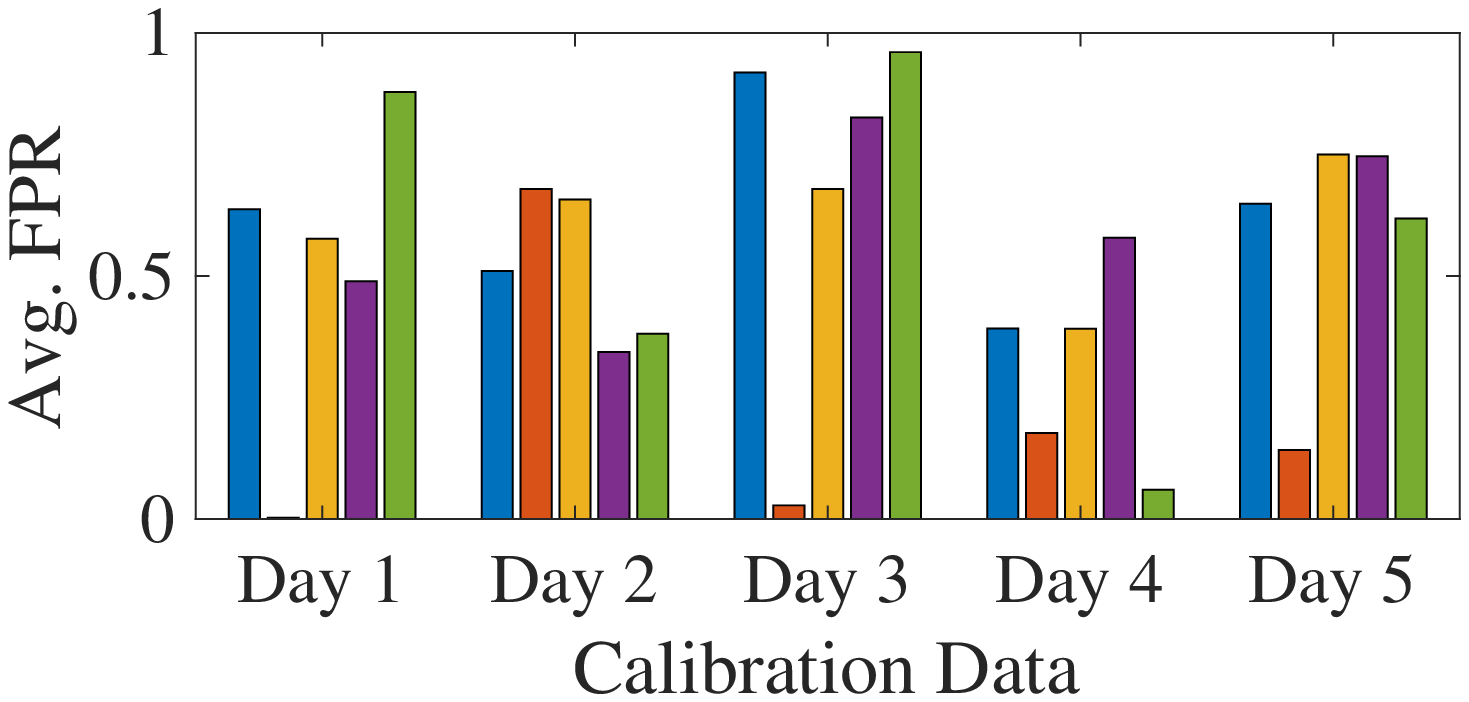}\label{subfig:day2_FPR}}
   }
\end{tabular}
\caption{\textbf{Channel Portability:} All models are trained on Day 1 data. $N$ = 10\%  and $M$ = 10.}
\label{fig:diff_days_open}
\end{figure}

\begin{figure*}
\centering
\begin{tabular}{m{0.65\columnwidth} m{0.25\columnwidth} m{0.65\columnwidth} m{0.25\columnwidth}}
\multirow{2}{*}[2em]{
\renewcommand{\thesubfigure}{a}
\subfloat[AUROC - Config. 1]{
   \includegraphics[width=0.65\columnwidth]{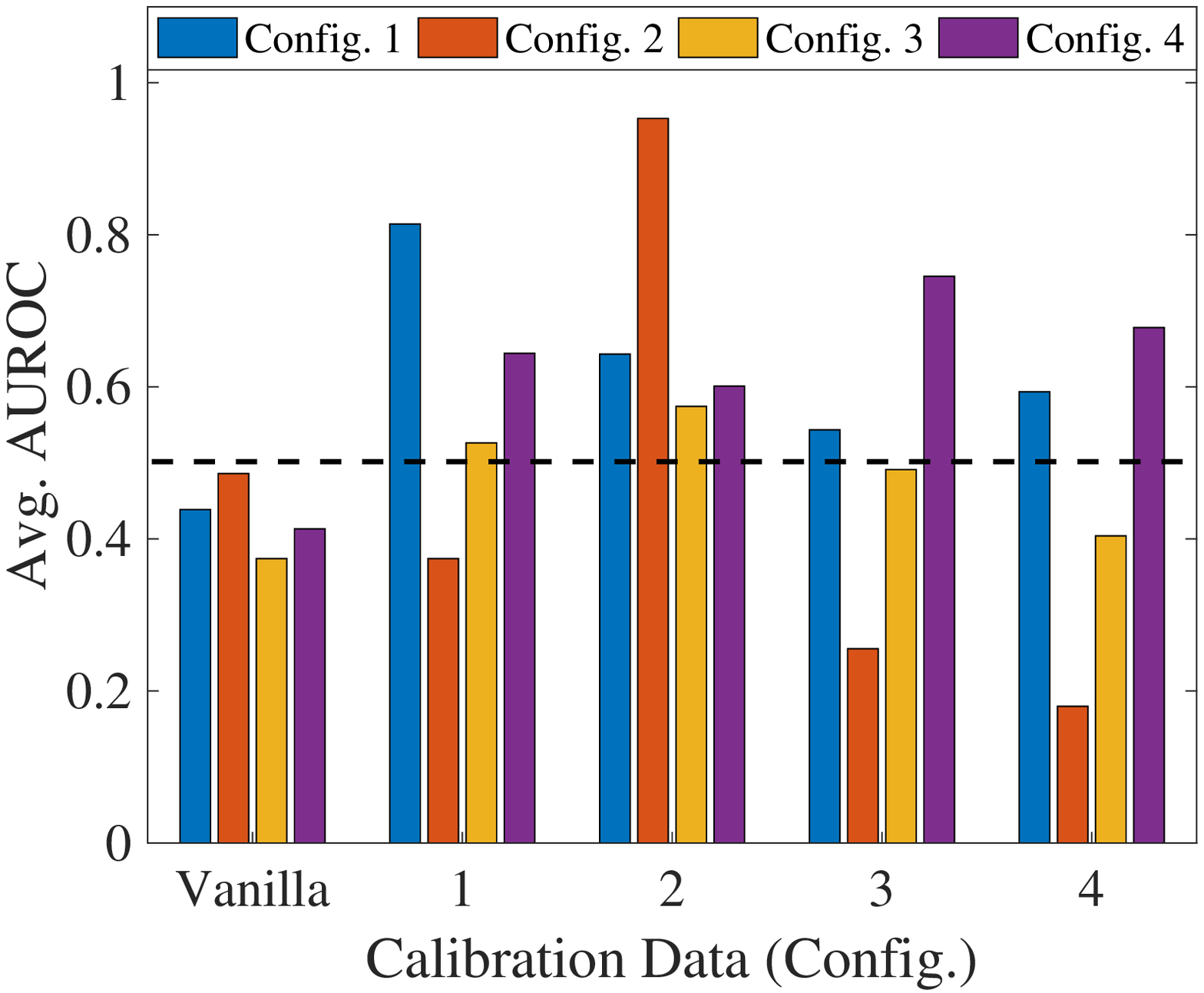}\label{subfig:config1_AUROC}}
   }
   &
\renewcommand{\thesubfigure}{b}
\subfloat[TPR - Config. 1]{
   \includegraphics[width=0.3\columnwidth]{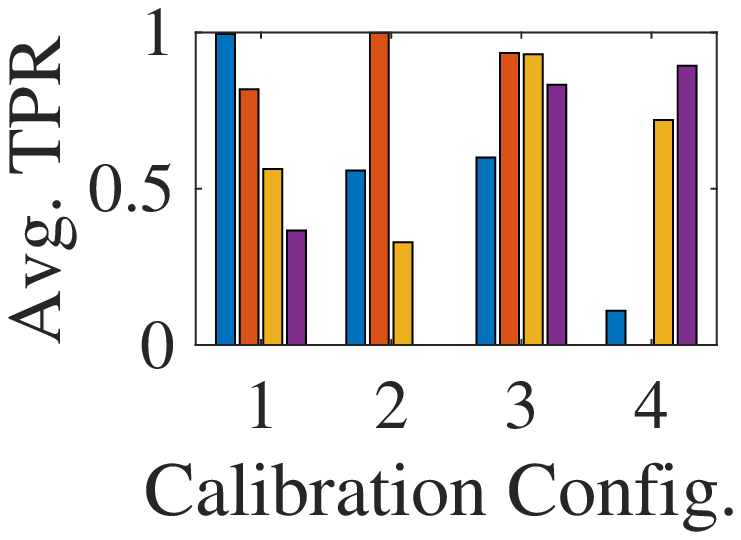}\label{subfig:config1_TPR}}
   &
   \multirow{2}{*}[2em]{
\renewcommand{\thesubfigure}{d}
\subfloat[AUROC - Config. 2]{
   \includegraphics[width=0.65\columnwidth]{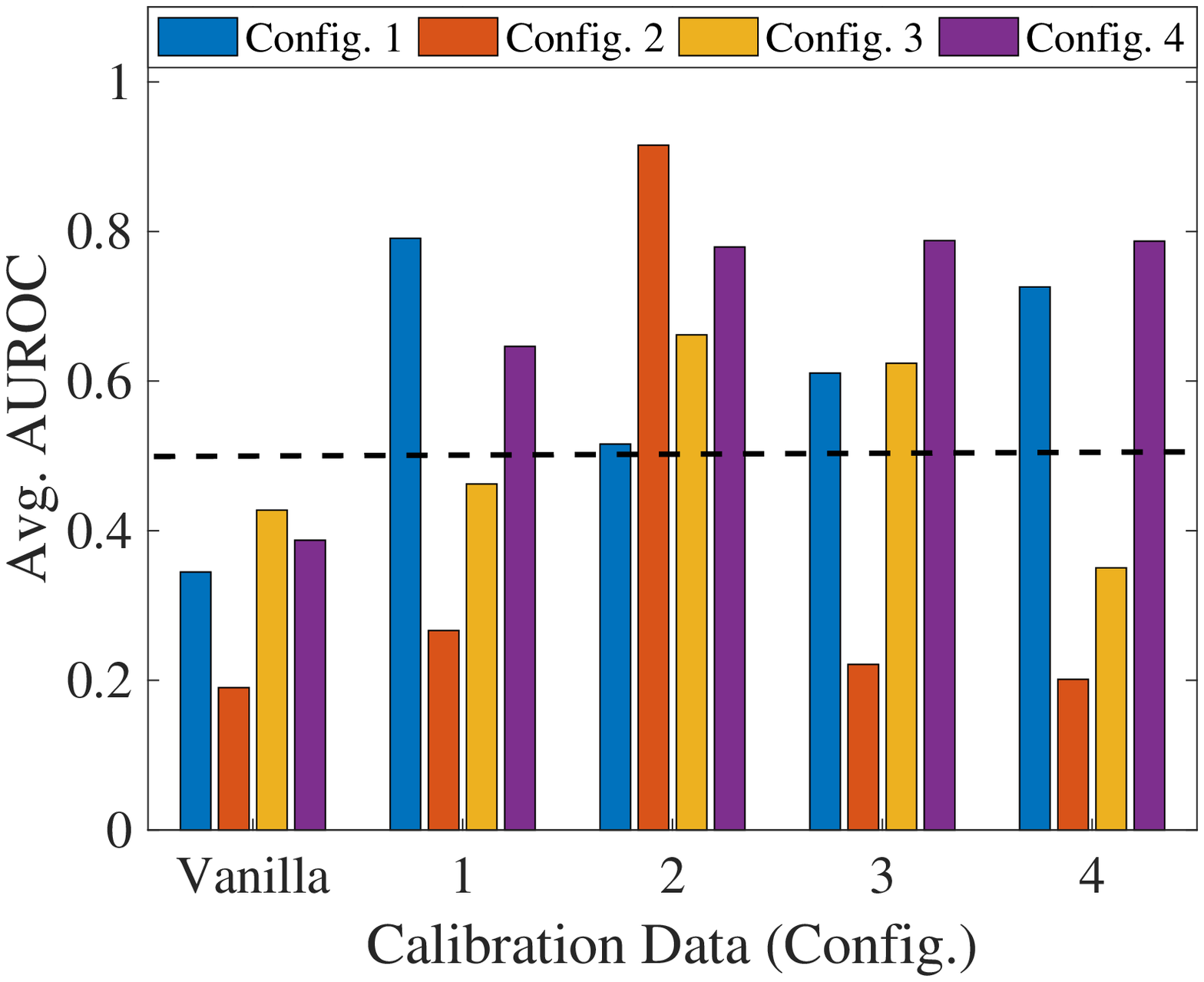}\label{subfig:config2_AUROC}}
   }
   &
\renewcommand{\thesubfigure}{e}
\subfloat[TPR - Config. 2]{
   \includegraphics[width=0.3\columnwidth]{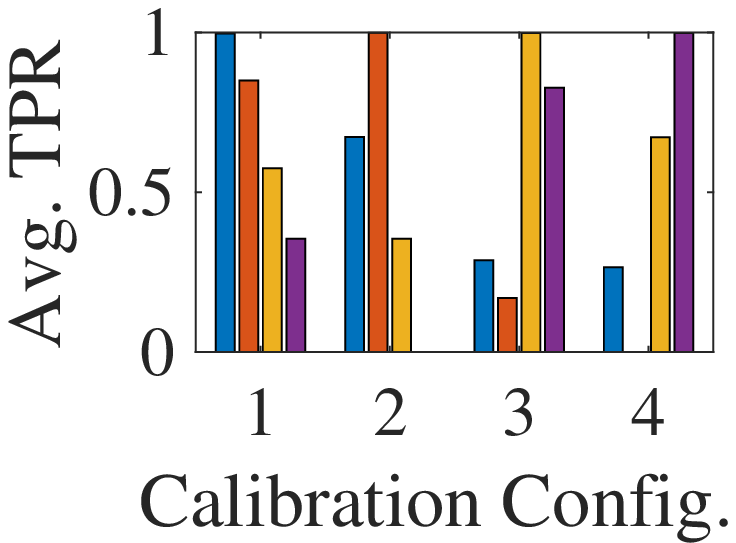}\label{subfig:config2_TPR}}
\\
   &
\renewcommand{\thesubfigure}{c}
\subfloat[FPR - Config. 1]{
   \includegraphics[width=0.3\columnwidth]{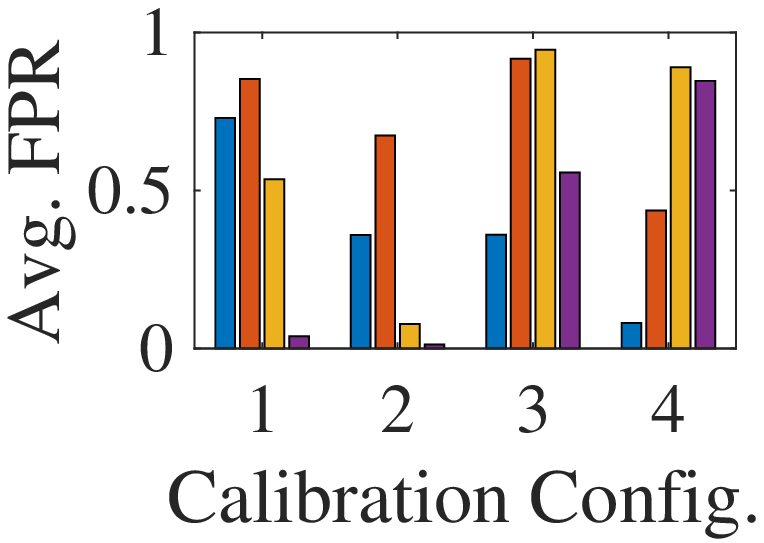}\label{subfig:config1_FPR}}
   &
   &
\renewcommand{\thesubfigure}{f}
\subfloat[FPR - Config. 2]{
   \includegraphics[width=0.3\columnwidth]{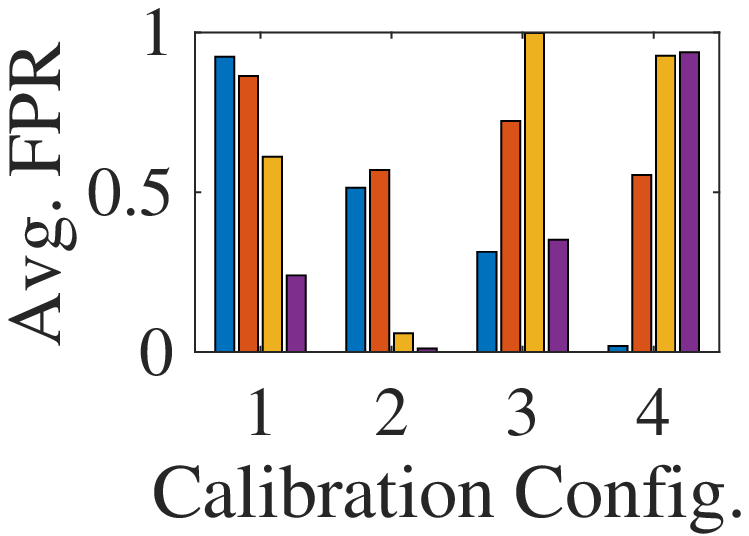}\label{subfig:config2_FPR}}
   \\
\end{tabular} 
\caption{\textbf{Configuration Portability}: Models trained on data using (a)-(c) Config. 1 and (d)-(f) Config. 2 and tested on data from transmitters using four different configurations. $N$ = 10\% and $M$ = 10.}
\label{fig:diff_configs_open}
\end{figure*}

\subsubsection{Channel Portability}
\proposed's portability across different channels is evaluated using channel portability dataset scenarios listed in Table~\ref{table:scenarios}. For this, \proposed\ was first trained on data collected on Day 1, and then calibrated for each of the five days. \proposed\ was tested with data from all five days.
The results are shown in Figs.~\ref{subfig:day1_AUROC}-\ref{subfig:day1_FPR}. Note that this evaluation was also performed for a model trained on Day 2 data, and similar results were achieved, but omitted here due to space limitations.

First considering the AUROC results in Fig.~\ref{subfig:day1_AUROC}, there are a couple of interesting trends that are worth mentioning. From observation of this figure it is evident that high AUROC is achieved when a model is calibrated with data collected on one day and tested using data from the same day. For instance, the model calibrated with Day 2 data performs very well when tested with Day 2 data. This trend exists regardless of the calibration day. The lowest Avg. AUROC achieved when models are calibrated/tested on data from the same day is $\sim$0.79. This trend is indicative of the ability of \proposed\ to achieve wireless channel portability for device authentication.

Another trend of note when looking at the AUROC results is that sometimes good performance is also achieved when testing using data from days other than the day used for calibration. For instance, in Fig.~\ref{subfig:day1_AUROC}, it can be seen that the model calibrated with Day 1 data also performs very well when tested with Day 2 data, and that the model calibrated with Day 4 data also performs very well when tested with Day 5 data. This trend is consistent, and is also present when the model is instead trained with data from Day 2. This could be indicative of some similarity between the wireless channel on these different days, but more work needs to be done to confirm this idea.

Moving to observe the TPR results in Figs.~\ref{subfig:day1_TPR}, it can be seen that the desired trend is present. That is, TPR is high when a model is calibrated and tested on data from the same day. This indicates that \proposed\ is excellent at correctly `admitting' examples that come from known devices.

Turning to the FPR results, it can be seen that the trend present in the `hardware portability' evaluation is again present here. That is, the FPR results are less than ideal, since calibrating and testing with data from the same day does not always produce the lowest FPR. This can be attributed to the idea explained in the `hardware portability' evaluation section.

\subsubsection{Configuration Portability}

To evaluate the configuration portability of \proposed, the configuration portability dataset in Table~\ref{table:scenarios} was used to train two twin network models: one with only data using Config. 1 and one with only data using Config. 2. Each of these trained models was then calibrated for each of the four configurations. Finally, \proposed\ was tested with data from all four configurations.

The results when training on data collected while the transmitters use Config. 1 are shown in Figs.~\ref{subfig:config1_AUROC}-\ref{subfig:config1_FPR} and those when the transmitters use Config. 2 are shown in Figs.~\ref{subfig:config2_AUROC}-\ref{subfig:config2_FPR}.
First observing the AUROC results, it can be seen that regardless of the training configuration, calibrating the model using Config. 1 or 2 results in the highest performance of the model occurring when it is tested with the same configuration as calibration (which is expected). One example of this can be seen in Fig.~\ref{subfig:config1_AUROC}, where the model calibrated with Config. 1 data performs the best when tested with Config. 1 data, and the model calibrated with Config. 2 data performs the best when tested with Config. 2 data. This trend is in line with the expected trends that have been mentioned thus far.

A unique trend present here can be seen by observing the performance of the models when calibrated and tested with Config. 3 data. In these cases, the performance of the model is not the best when testing on the same domain as calibration. In particular, testing with Config. 4 produces better performance than testing with Config. 3. This is not the case however, when the model is trained with data from transmitters that use Config. 3 or Config. 4 (not pictured). In this case, the result is as expected, where calibrating and testing with Config. 3 data produces the best result of all tests for that calibration. This result could be attributed to the fact that generally Configs. 1 and 2 are more similar to each other and less similar to Configs. 3 and 4 (see Table~\ref{table:lora}).

The TPR results for the models trained with either configuration are consistent with the desired trend. That is, a very high TPR is achieved when the model is calibrated and tested using data from the same configuration. 

Similar to the other evaluations, the FPR results are less encouraging than those for AUROC and TPR. There are many instances where calibrating and testing using data from the same configuration results in the highest FPR among all of the tests for that calibration. This can be explained using the same line of thinking used in the `hardware portability' evaluation. That is, achieving a low FPR is easier when calibration and testing are done with data from two different configurations. This conclusion is reinforced by the FPR results here, as generally the FPR is lower when testing on a configuration that is `more different' from the one used for calibration.

\begin{figure}
\centering
\begin{tabular}{m{0.55\columnwidth} m{0.4\columnwidth}}
\centering
\multirow{2}{*}[2em]{
\subfloat[AUROC]{
   \includegraphics[width=0.58\columnwidth]{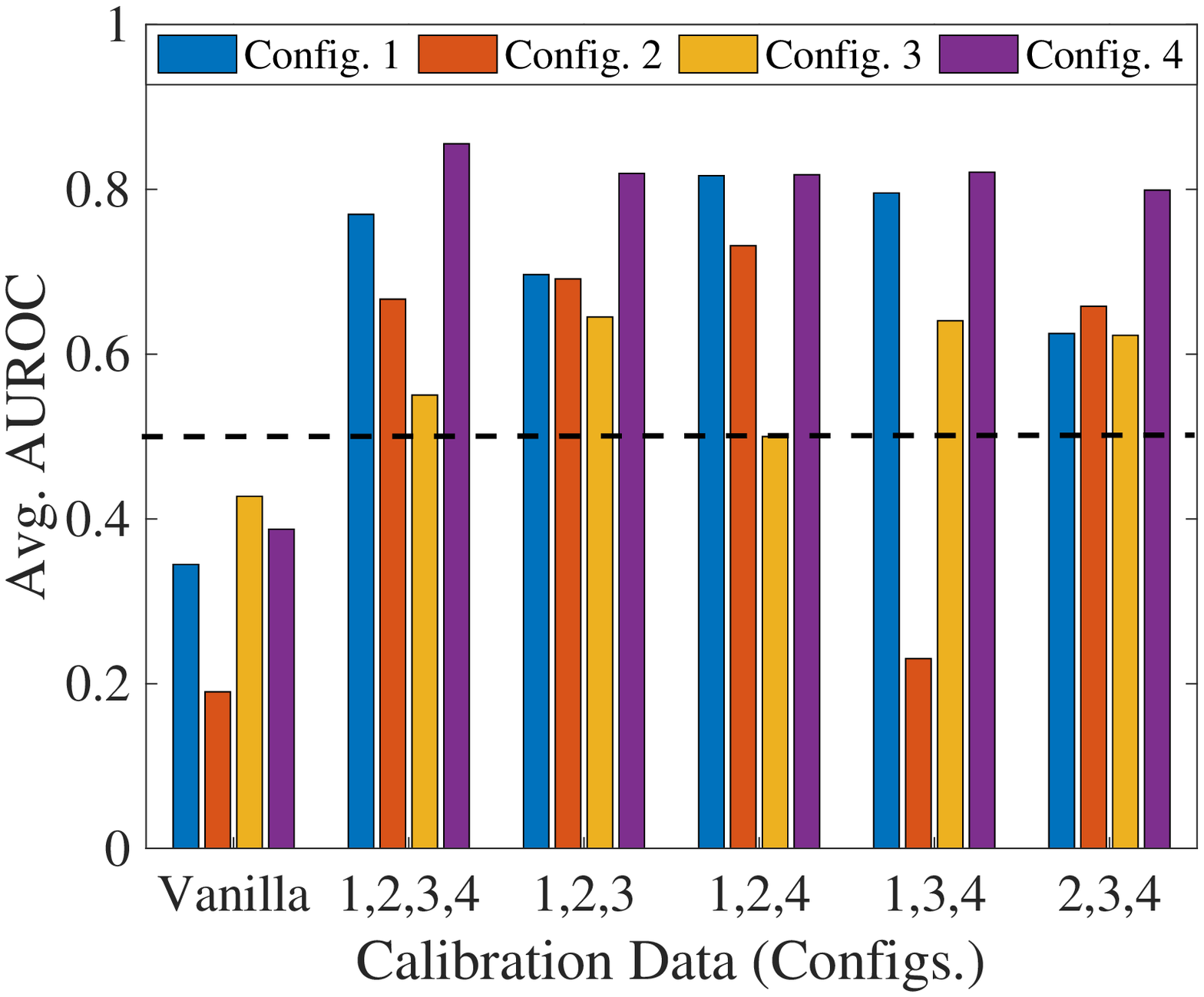}\label{subfig:config2_multi_AUROC}}
   }
   &
\subfloat[TPR]{
   \includegraphics[width=0.40\columnwidth]{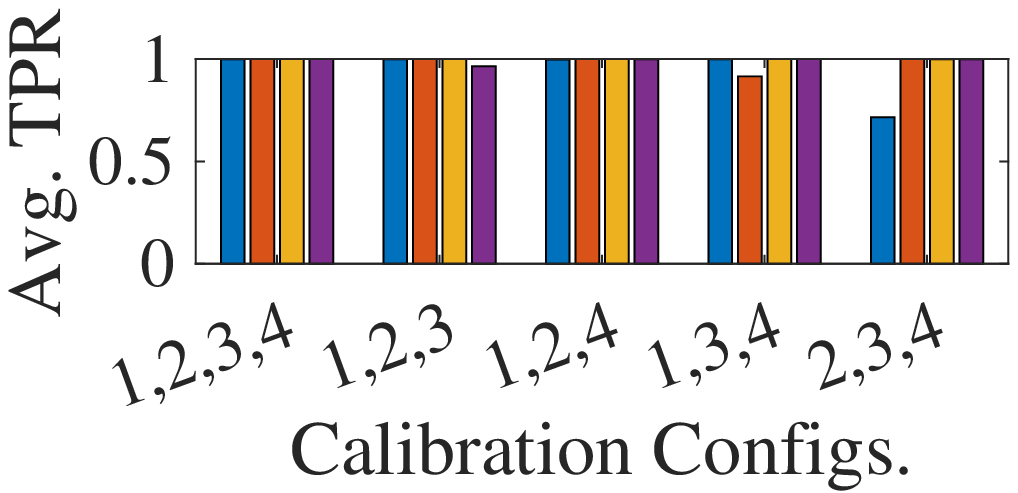}\label{subfig:config2_multi_TPR}}
   \\
   &
\subfloat[FPR]{
   \includegraphics[width=0.40\columnwidth]{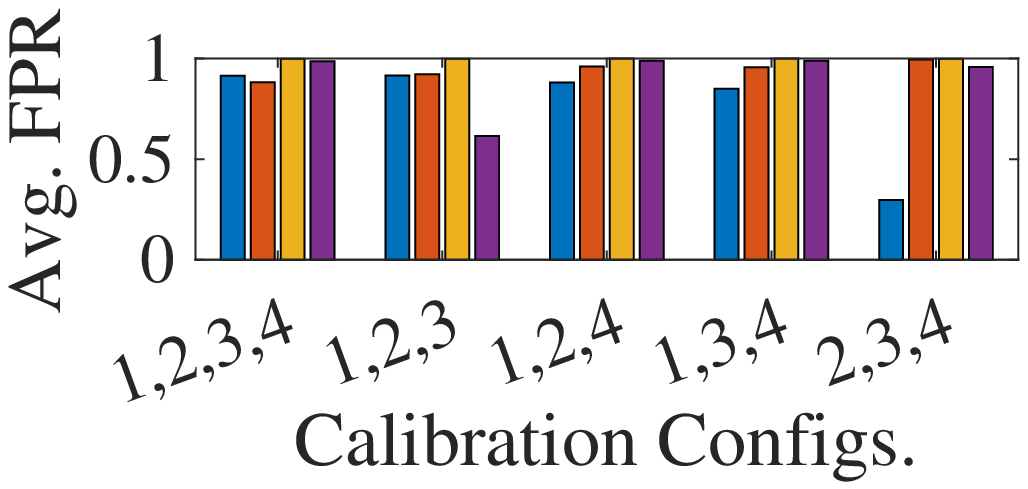}\label{subfig:config2_multi_FPR}}
   \\
\end{tabular}
\caption{\textbf{Multi-Calibration Configuration Portability:} Models trained on data using Config. 2 and calibrated with data from multiple configurations. $N$ = 10\% and $M$ = 10.}
\label{fig:diff_configs_multi_cal_open}
\end{figure}

Finally, the last evaluation performed in this section leverages the notion that \proposed\ could be used to calibrate for more than one configuration by performing multiple consecutive calibrations. To this end, the twin network model trained using Config. 2 data was calibrated using data from multiple configurations and tested on data from all four configurations. The results of these tests are shown in Fig.~\ref{fig:diff_configs_multi_cal_open}. The results here are encouraging for AUROC and TPR, but contain the same issues mentioned above when it comes to FPR results. There is also a notable drop in AUROC when calibrating for multiple configurations when compared to calibrating for a single configuration.

\section{Conclusion}
\label{sec:conc}
This work proposes \proposed, a technique for creating a portable deep neural network for LoRa device authentication which can be calibrated quickly, using a small amount of labeled examples, to perform effectively on data from a domain other than the one it was originally trained with. It has been demonstrated through experimentation with a testbed of IoT devices that \proposed\ enables portability with respect to receiver hardware, a change in the wireless channel due to the passage of time, or a change in the configuration of the transmitters.

\bibliographystyle{IEEEtran}
\bibliography{References}

\end{document}